\documentclass[
]{journal}

\usepackage{orcidlink,thumbpdf,lmodern}

\usepackage[utf8]{inputenc}

\author{
Paul Murrell\\The University of Auckland
}
\title{Data Verbalisation: What is Text Doing in a Data Visualisation?}

\Plainauthor{Paul Murrell}
\Plaintitle{Data Verbalisation: What is Text Doing in a Data Visualisation?}
\Shorttitle{Data Verbalisation}

\Abstract{
This article discusses the role that text elements play in a data visualisation. We argue that there is a need for a simple, coherent explanation of text elements similar to the understanding that already exists for non-text elements like bars, points, and lines. We explore examples of how text is used within a data visualisation and use existing knowledge and assessment techniques to evaluate when text is effective and when it is not. The result is a framework that aims to be easy to understand and easy to apply in order to understand the purpose and effectiveness of the text elements in any data visualisation.
}

\Plainkeywords{data visualisation, text}

\Address{
    Paul Murrell\\
    The University of Auckland\\
    Department of Statistics\\
The University of Auckland\\
Private Bag 92019\\
Auckland 1010\\
New Zealand\\
  E-mail: \email{paul@stat.auckland.ac.nz}\\
  URL: \url{https://www.stat.auckland.ac.nz/~paul/}\\~\\
  }

\usepackage{longtable,booktabs,array}
\usepackage{calc} 
\usepackage{etoolbox}
\makeatletter
\patchcmd\longtable{\par}{\if@noskipsec\mbox{}\fi\par}{}{}
\makeatother
\IfFileExists{footnotehyper.sty}{\usepackage{footnotehyper}}{\usepackage{footnote}}
\makesavenoteenv{longtable}

\usepackage{booktabs}
\usepackage{longtable}
\usepackage{array}
\usepackage{multirow}
\usepackage{wrapfig}
\usepackage{float}
\usepackage{colortbl}
\usepackage{pdflscape}
\usepackage{tabu}
\usepackage{threeparttable}
\usepackage{threeparttablex}
\usepackage[normalem]{ulem}
\usepackage{makecell}
\usepackage{xcolor}

\usepackage{amsmath} \usepackage{booktabs} \usepackage{paralist} \usepackage{dirtytalk} \usepackage[width=.48\hsize]{subfig}

\begin{document}

\newcommand{\class}[1]{`\code{#1}'}
\newcommand{\fct}[1]{\code{#1()}}
\newcommand{\ma}[1]{\ensuremath{\mathbf{#1}}}

\hypertarget{sec:intro}{%
\section{Introduction}\label{sec:intro}}

Figure \ref{fig:drama} shows two different presentations of data on the
number of youth offenders in New Zealand over three years, broken down
by sex, age group, and type of offence: a multi-panelled bar plot and
a block of text (1,000 words long).
This is a classic demonstration of the value of data visualisation.
Although the same information is available within the two formats,
many useful features, such as the difference between males and females,
are much more easily and rapidly extracted from the bar plots
compared to the block of text.
Although the block of text contains statements like
``theft is the most common offence'' and ``Comparing the data from 2022 to 2024
\ldots{} theft consistently remains the most common offence'',
these features are both easier and faster to extract from the bar plot.

\begin{CodeChunk}
\begin{figure}

{\centering \includegraphics[width=0.7\linewidth]{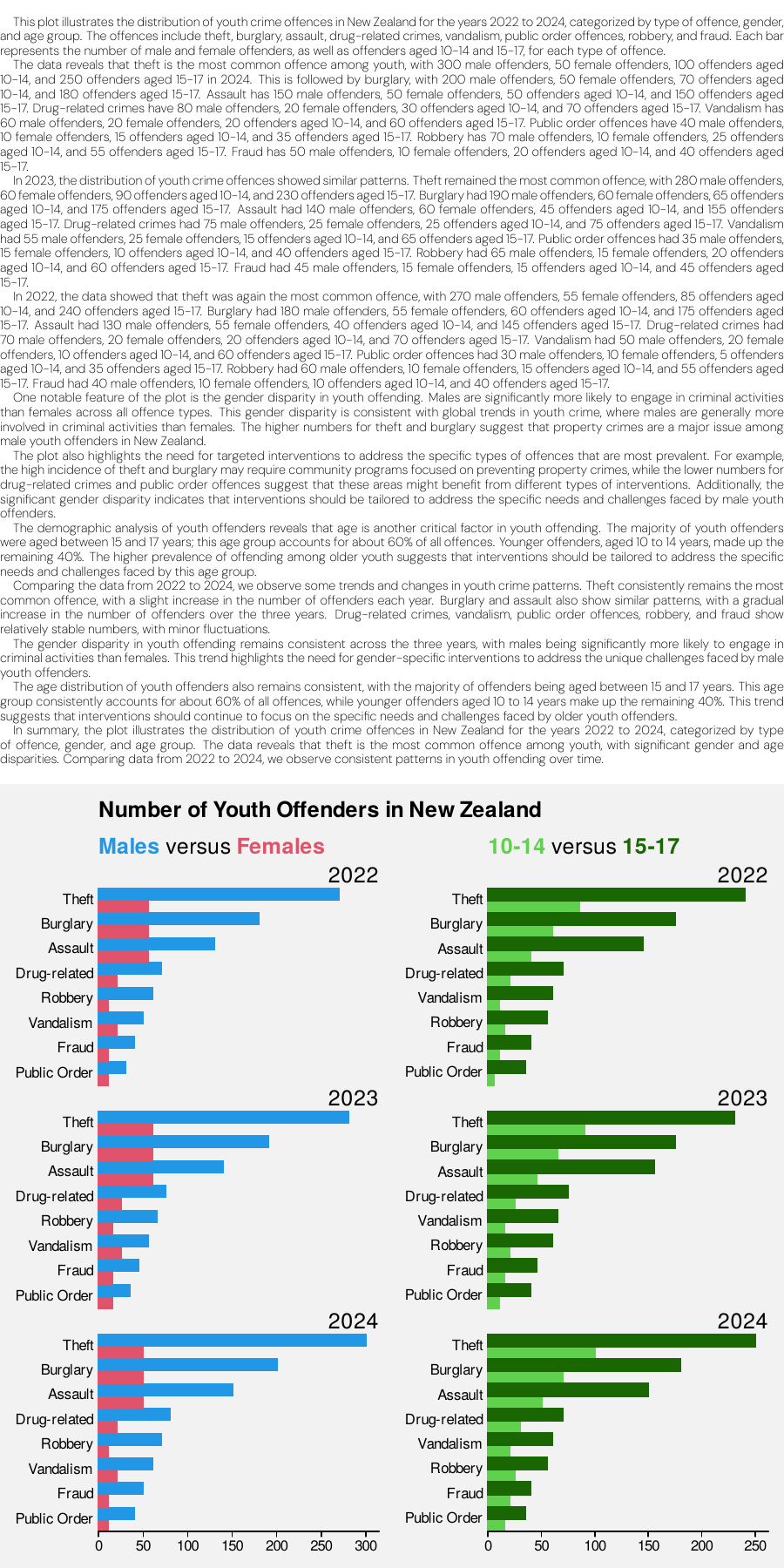} 

}

\caption[Two presentations of the number of youth offenders in New Zealand from 2022 to 2024, broken down by sex, age group, and type of offence]{Two presentations of the number of youth offenders in New Zealand from 2022 to 2024, broken down by sex, age group, and type of offence: A multi-panelled bar plot (bottom) and 1,000 words (top). It is much easier and faster to extract features and trends from the bar plot.}\label{fig:drama}
\end{figure}
\end{CodeChunk}

Table \ref{tab:table} shows another, almost entirely text-based
presentation of the data on
the number of youth offenders. Again, it is much harder to
extract features, such as trends over time,
from this table compared to the bar plots.

\begin{CodeChunk}
\begin{table}
\centering
\caption{\label{tab:table}A tabular presentation of the number of youth offenders in New Zealand from 2022 to 2024, broken down by sex, age group, and type of offence. It requires more time and effort to extract features and trends from this table compared to the bar plot in Figure \ref{fig:drama}.}
\centering
\fontsize{7}{9}\selectfont
\begin{tabular}[t]{rlrrrr}
\toprule
\textbf{Year} & \textbf{Offence} & \textbf{Male} & \textbf{Female} & \textbf{10-14} & \textbf{15-17}\\
\midrule
2022 & Theft & 270 & 55 & 85 & 240\\
2022 & Burglary & 180 & 55 & 60 & 175\\
2022 & Assault & 130 & 55 & 40 & 145\\
2022 & Drug-related & 70 & 20 & 20 & 70\\
2022 & Robbery & 60 & 10 & 15 & 55\\
2022 & Vandalism & 50 & 20 & 10 & 60\\
2022 & Fraud & 40 & 10 & 10 & 40\\
2022 & Public Order & 30 & 10 & 5 & 35\\
\addlinespace
2023 & Theft & 280 & 60 & 90 & 230\\
2023 & Burglary & 190 & 60 & 65 & 175\\
2023 & Assault & 140 & 60 & 45 & 155\\
2023 & Drug-related & 75 & 25 & 25 & 75\\
2023 & Robbery & 65 & 15 & 20 & 60\\
2023 & Vandalism & 55 & 25 & 15 & 65\\
2023 & Fraud & 45 & 15 & 15 & 45\\
2023 & Public Order & 35 & 15 & 10 & 40\\
\addlinespace
2024 & Theft & 300 & 50 & 100 & 250\\
2024 & Burglary & 200 & 50 & 70 & 180\\
2024 & Assault & 150 & 50 & 50 & 150\\
2024 & Drug-related & 80 & 20 & 30 & 70\\
2024 & Robbery & 70 & 10 & 25 & 55\\
2024 & Vandalism & 60 & 20 & 20 & 60\\
2024 & Fraud & 50 & 10 & 20 & 40\\
2024 & Public Order & 40 & 10 & 15 & 35\\
\bottomrule
\end{tabular}
\end{table}

\end{CodeChunk}

We could naively conclude from
Figure \ref{fig:drama} and Table \ref{tab:table}
that text-based presentations of data values
are bad and should be avoided in favour of more geometric data symbols,
like bars. However, the situation is clearly
not that simple because the multi-panel bar plot in Figure
\ref{fig:drama} contains not just bars, but also text elements, including
titles and axis labels.
Furthermore, Figure \ref{fig:notext}
shows, by removing all of the text elements from Figure \ref{fig:drama},
that the
text elements in the multi-panel bar plot are serving an important
purpose. We can see that some values are larger than another values in
Figure \ref{fig:notext}, but we cannot tell what the values are nor
how large the differences are.

\begin{CodeChunk}
\begin{figure}

{\centering \includegraphics[width=0.7\linewidth]{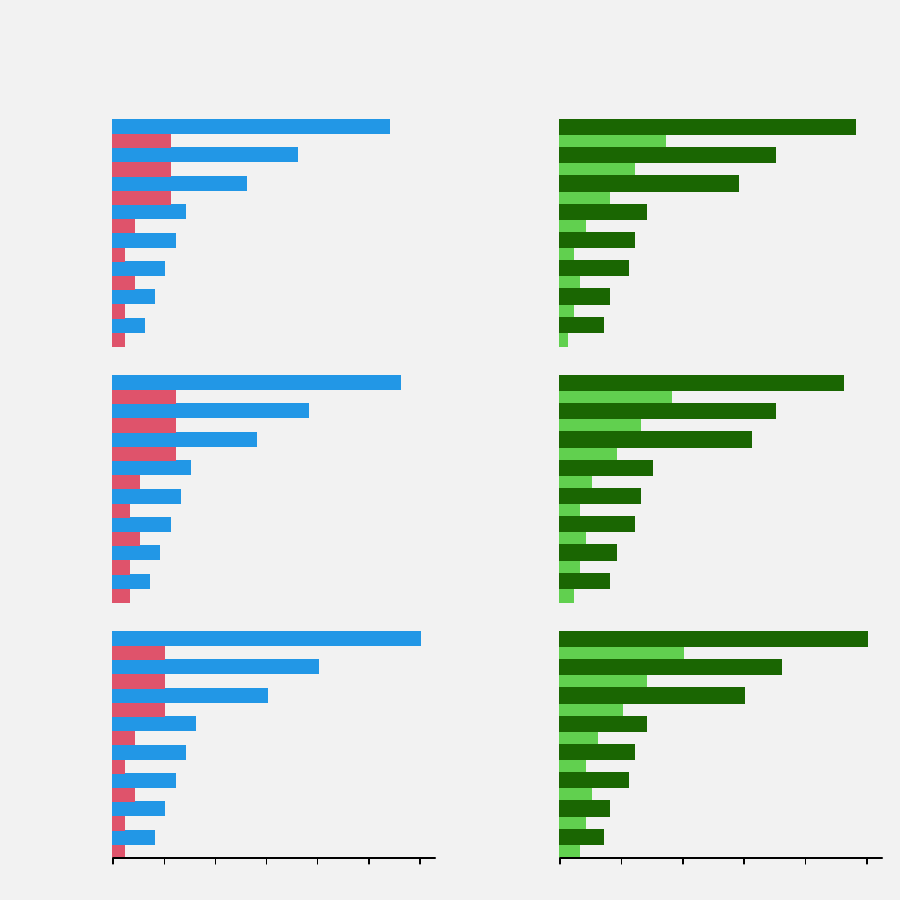} 

}

\caption[A multi-panelled bar plot showing the number of youth offenders
in New Zealand from 2022 to 2024, broken down by sex, age group, and type of
offence. This is the bar plot from Figure \ref{fig:drama} with all of the text
elements removed. Although it is possible to perceive relative trends, we cannot tell what the different colours represent nor what absolute amounts are represented by a particular bar length.]{A multi-panelled bar plot showing the number of youth offenders
in New Zealand from 2022 to 2024, broken down by sex, age group, and type of
offence. This is the bar plot from Figure \ref{fig:drama} with all of the text
elements removed. Although it is possible to perceive relative trends, we cannot tell what the different colours represent nor what absolute amounts are represented by a particular bar length.}\label{fig:notext}
\end{figure}
\end{CodeChunk}

It is clear that a bar plot is more effective at conveying data features
than a table of text and clearer than paragraphs
of text, but it is also clear that a bar plot must itself contain text elements.
The goal of this article is to discuss what role text elements play within
a data visualisation. What are the strengths and weaknesses of text
elements and, consequently, when they should be used and when they
should be avoided.

\subsection{Text in the data visualisation literature}

There are many publications that provide general guidelines for
producing effective
data visualisations.
These usually contain advice about the text labels within a
data visualisation, including useful guidelines regarding
the \emph{content} of labels, such as the information that should be
included in titles.
For example, \cite{cleveland1994elements} provided the following three
guidelines for captions and titles:

\begin{compactenum}
\item \say{Describe everything that is graphed.}
\item \say{Draw attention to the important features of the data.}
\item \say{Describe the conclusions that are drawn from the data on the graph.}
\end{compactenum}

\cite{tufte2001visual} proposes principles for labels that include:
\say{Clear, detailed, and thorough labeling should be used to defeat graphical
distortion and ambiguity} and \say{Label important events in the data.}
More recently, \cite{Wilke2019FDV} and \cite{Wickham2023R} exhort us to
``summarize the main finding'', ``describe the source of the data'', and
``state \dots the units in which the variables are measured''.
\cite{schwabish2021better} suggests that we
``write the title like a newspaper headline''.

While these guidelines provide useful advice about the content of text labels,
particularly titles and captions,
there is an assumption about what text will be included on a plot,
for example that all
plots will have a title,
coupled with a silence about other text elements, such as axis labels.
There is an emphasis on what to do, but less of an explanation of why
we should do it \citep{stokes2022textoftenbetterthemes}.
For example, in Figure \ref{fig:drama}, is there a good reason
for having two sub-titles as well as a main title?
How do we decide on an effective placement of the sub-titles?

Text is also often treated as a special type of element
that is distinct from other elements
within a data
visualisation.
A common breakdown of a data visualisation identifies
the title, axes, and legends as separate components, with text
featuring prominently in those, as distinct from data symbols, which are used to
represent the data values \citep[e.g.,][]{cleveland1994elements}.
\cite{cairo2019how}
distinguishes between
the ``scaffolding'' and the ``content'' of a data visualisation,
with the former consisting of titles, scales, legends, and annotations.
\cite{28d110e0-3c98-3aca-8c91-75f5aed6a481} separates
forms of science communication that are based on text from
forms that are based on graphical symbols.

This separation encourages us to think of text as a completely
different type of component
within a data visualisation and ignores the
commonalities between text and other elements. For example, how
do we explain the fact that colour
is simultaneously
used effectively in both the titles and the bars in Figure \ref{fig:drama}?

More theoretical publications on data visualisation
provide explanations for why certain representations
of data values are more effective than others.
These works explain why, for example,
bars in a bar plot are effective representations of quantitative data values
and superior to using wedges in a pie chart.
However, there is often no corresponding explanation of when or why
text can also be effective in a data visualisation.
For example, \cite{wilkinson2005grammar}
does not include text as a ``geometry'' alongside points, lines, and areas
and includes only ``label'' as a special aesthetic that is specific to text.
\cite{ware2020information} distinguishes between
``sensory'' symbols, which are perceived automatically, like bars,
and ``arbitrary'' symbols, like text, that require learning and
focuses more on the effectiveness of sensory symbols.
Text is mostly treated as informative labels and annotations,
rather than as effective representations of data values.
\cite{munzner2014visualization} does not include text as a ``mark''
or in relation to ``visual channels'' used to
represent data values.
\cite{bertin1983semiology} also does not consider text as a graphical mark
\citep[text only makes a very brief, 
untranslated, appearance;][]{Brath04032019}.

\subsection{Integrating text}

The problem that this article attempts to solve is the fact that
an explanation for how text works in a data visualisation is
incomplete, or at least is more fragmented and less coherent
than the existing explanations for non-text elements in a data visualisation
\citep{10.1145/3593580}.
In brief,
the solution that this article proposes is that we view text elements
through the same lense that we use
to view other elements of a data visualisation,
like the bars in a bar plot and the points in a scatter plot,
and that we assess text elements using the same
criteria that we use to assess the other elements of a data visualisation,
like perceptual accuracy.
If we consider
a data visualisation to be an encoding of data values as
the visual features of data symbols, for example,
counts as the heights of bars, then
text is just another data symbol with visual features that we can
use to encode data values.
The effectiveness of text elements within a data visualisation
will then depend on the appropriateness
of the encoding from data values to
text visual features and the effectiveness of the decoding back to data values.

This article draws together existing work from many places,
including many of the works referenced in the previous section.
We will make particular reference to \cite{brath2020visualizing},
which is a significant exception within the data visualisation literature
in that it explicitly integrates text
alongside more traditional representations of data values such
as points, lines, and bars, following on from
\cite{persee.fr:colan_0336-1500_1980_num_45_1_1369}.
Although that work is very thorough, and we will assimilate several
ideas from that work,
this article
takes a slightly different approach to integrating text.
A more detailed comparison with
\cite{brath2020visualizing} is provided in Appendix \ref{app:a}.
Another goal of this
article is to bring some of these ideas to the attention of
a statistics audience that might be less familiar with the 
work on text elements that is distributed throughout the broader
data visualisation literature.
The contribution of this article is to synthesise many of
those works and to provide some new insights within
a single, relatively simple, and coherent description of how text works
in a data visualisation.

We will view text as just another data symbol
with visual features that we can use to encode data values.
By integrating text within the same framework that we use for
other data symbols, like bars and points, and other visual features,
like length, position, and colour, we will obtain a deeper understanding
of when and why to use text within a
data visualisation.

In Section \ref{sec:framework} we will briefly review the established
framework for understanding how non-text elements work within a
data visualisation. Sections \ref{sec:chars} to \ref{sec:effectiveness}
will explore the use of text as data symbols within standard
plots, with a direct comparison to non-text data symbols.
We will see that in many ways text works just like other data symbols,
but also that text has its own special advantages and disadvantages.
Sections \ref{sec:phrases} and \ref{sec:mem} will look at the use of
larger blocks of text elements within plots and show that
text is uniquely appropriate for those roles.
Section \ref{sec:tables} will revisit the presentation of data in
tables in the light of the previous sections and will encourage
less of a two-worlds view of text versus non-text symbols.

The framework that we will develop is summarised in
Section \ref{sec:summary}. This framework is a synthesis and
a simplification of many other works,
but the hope is that the result is easy to comprehend
and easy to apply.
Appendix \ref{app:a}
acknowledges some of the details that were
glossed over in the development of the framework,
and some of the limitations that result, but also
where this framework makes its own contributions.

\newpage

\hypertarget{sec:framework}{%
\section{Encoding data values as data symbols}\label{sec:framework}}

We begin with a brief review of the standard framework
for explaining how non-text data symbols work in a
data visualisation.

A data visualisation can be characterised as
an \emph{encoding} of data values as the \emph{visual features} of
\emph{data symbols}. Different authors have used a variety of terms,
such as mappings instead of encodings \citep{wilkinson2005grammar},
data attributes instead of data values \citep{bertin1983semiology},
visual channels instead of visual features, and
marks instead of data symbols \citep{munzner2014visualization},
but the basic framework is the same.
For example, in the bar plot in Figure \ref{fig:drama},
the number of youth offenders (quantitative data value) is
be encoded as the length (visual feature) of a bar (data symbol)
and the sex of the offender (qualitative data value) is
encoded as the colour (visual feature) of a bar (data symbol).

The encoding from data values to visual features requires a \emph{scale},
which is a transformation from different data values to different values of the
visual feature \citep{wilkinson2005grammar}.
For example, in Figure \ref{fig:drama},
each value of sex is associated with
a different colour.

\begin{CodeChunk}


\begin{flushleft}\includegraphics[width=0.4\linewidth]{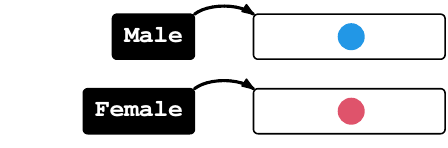} \end{flushleft}

\end{CodeChunk}

For a visual feature like length, each data value is ultimately associated
with a physical dimension (a number of centimetres or a number of pixels).
For example, in Figure \ref{fig:drama}, a count of zero corresponds
to a bar of length zero, and a count of 100 corresponds to a bar twice
as long as a bar representing a count of 50.
The physical quantities shown below (and in similar examples later on) are
indicative only.

\begin{CodeChunk}


\begin{flushleft}\includegraphics[width=0.4\linewidth]{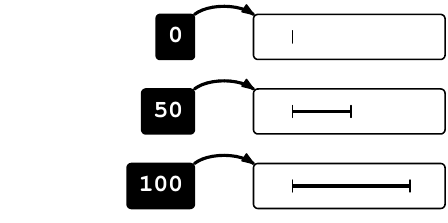} \end{flushleft}

\end{CodeChunk}

As another example, Figure \ref{fig:rwc}\subref*{fig:rwc-1}
shows a scatter plot of the number of points scored
versus the number of points conceded,
both quantitative measures, for teams at the 2023 Rugby World Cup.
Here, the data symbol is a data point (a small circle) and the quantitative
data values are encoded as the horizontal and vertical positions of
the points. Qualitative data values, the hemisphere for each country,
are encoded as the colours of the points.

\begin{CodeChunk}


\begin{flushleft}\includegraphics[width=0.4\linewidth]{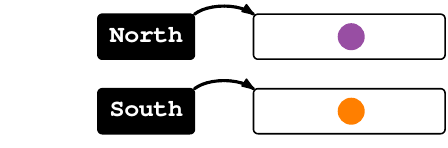} \includegraphics[width=0.4\linewidth]{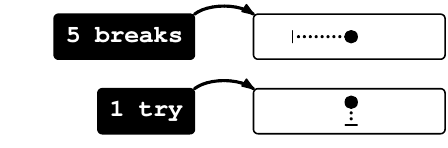} \end{flushleft}

\end{CodeChunk}

\begin{CodeChunk}
\begin{figure}

{\centering \subfloat[The data symbols are data points. We can decode points scored and points conceded from the positions of the data points. We can also decode which hemisphere a team is from based on the colours of the data points\label{fig:rwc-1}]{\includegraphics[width=0.49\linewidth]{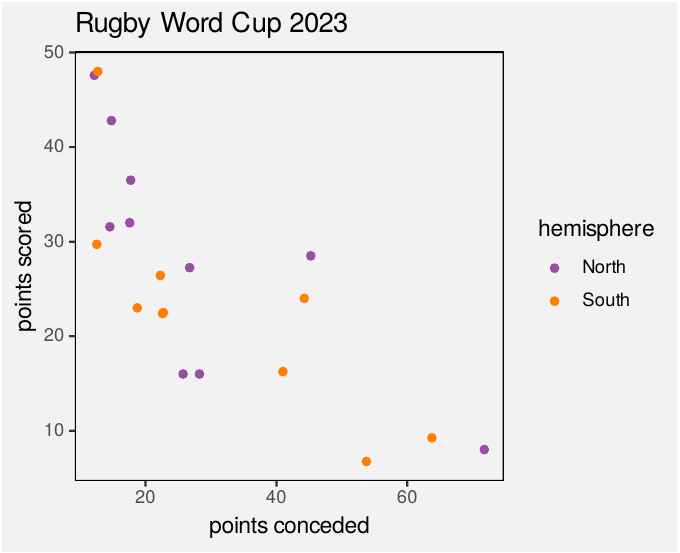} }\subfloat[The data symbols are text characters.  We can decode points scored and points conceded from the positions of the data points. We can also decode which hemisphere a team is from based on the shapes of the characters.\label{fig:rwc-2}]{\includegraphics[width=0.49\linewidth]{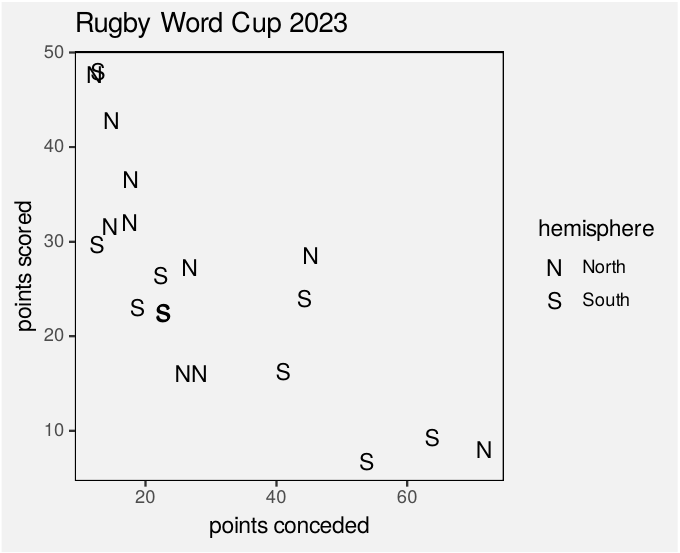} }

}

\caption[Scatter plots of the number of points scored and the number of points conceded (both are per-game averages) for teams in the 2023 Rugby World Cup]{Scatter plots of the number of points scored and the number of points conceded (both are per-game averages) for teams in the 2023 Rugby World Cup.}\label{fig:rwc}
\end{figure}
\end{CodeChunk}

Both the bar plot in Figure \ref{fig:drama} and the scatter plot in
Figure \ref{fig:rwc}\subref*{fig:rwc-1}
are effective because the data values are encoded
as visual features that can be \emph{decoded} effectively. We can
decode from the visual
features back to the data values.
For example, the lengths of bars and the positions of points
are effective visual features for decoding quantitative values, like
the number of offenders, the number of points scored, or the number of
points conceded
\citep{cleveland1984graphical, 10.1145/1753326.1753357}.
Colour is an effective visual feature for decoding qualitative values,
like the sex of the offender or the hemisphere for a country
\citep{munzner2014visualization}.

\begin{CodeChunk}


\begin{flushleft}\includegraphics[width=0.4\linewidth]{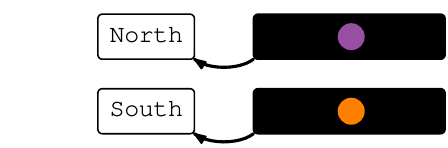} \includegraphics[width=0.4\linewidth]{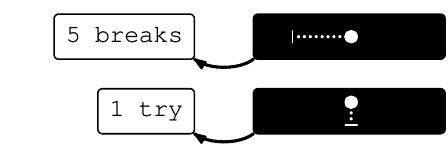} \end{flushleft}

\end{CodeChunk}

\hypertarget{sec:chars}{%
\section{Characters as data symbols}\label{sec:chars}}

We now begin to consider how to integrate text data symbols into this
standard framework.

What happens if we use text as the data symbols instead of points or bars?
The first thing to note is that text comes in a number of different forms.
For example, we can use a single character or we can use whole words, we
can use phrases or we can use
sentences, and so on \citep{brath2020visualizing}.
We will look first at using a single text character as a data symbol.

Figure \ref{fig:rwc}\subref*{fig:rwc-2}
shows a version of the scatter plot in
Figure \ref{fig:rwc}\subref*{fig:rwc-1}
that uses a text
character as the data symbol---an ``N'' for teams from the \texttt{North}
and an ``S'' for teams from the \texttt{South}.
One change from the original scatter plot in \ref{fig:rwc}\subref*{fig:rwc-1}
is the data symbol---a data point has become a text character---but
the encoding of quantitative values remains the same.
The number of points scored and the number of points conceded are encoded as
horizontal and vertical positions.

This tells us something immediately: Text data symbols share some
of the same visual features as other data symbols.
For example, we can encode a quantitative data value as the
position of a text character, just as we can encode a quantitative
data value as the position of a data point.

However, the encoding of the qualitative values in
Figure \ref{fig:rwc}\subref*{fig:rwc-2} is different
from the encoding in Figure \ref{fig:rwc}\subref*{fig:rwc-1}.
The different values of hemisphere
are encoded as different characters, rather than as different colours.
As with any encoding, there is a scale: each value of
hemisphere maps to a different text character.
Different characters have different visual appearances, so different
characters can be used to represent different categories of a
qualitative variable.

\begin{CodeChunk}


\begin{flushleft}\includegraphics[width=0.4\linewidth]{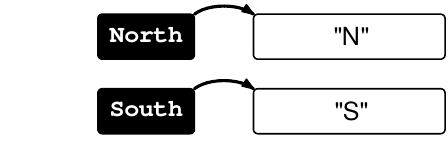} \end{flushleft}

\end{CodeChunk}

In Figure \ref{fig:rwc}\subref*{fig:rwc-2},
by encoding each hemisphere as a different
character, we are encoding the hemisphere
as the \emph{shape} of the text.
This is just like encoding qualitative
values using the shape visual feature
of data points in a scatter plot.
In other words, shape is another visual feature that
text shares with other data symbols.
Figure \ref{fig:textaes}\subref*{fig:textaes-1}
shows another version of the scatter plot
with data points as the data symbols and the hemisphere of each
country encoded as the shape of the data points.

\begin{CodeChunk}


\begin{flushleft}\includegraphics[width=0.4\linewidth]{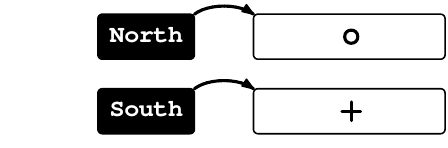} \end{flushleft}

\end{CodeChunk}

The only difference between
Figure \ref{fig:textaes}\subref*{fig:textaes-1}
and
Figure \ref{fig:rwc}\subref*{fig:rwc-2}
is the data symbol: text characters instead of data points.
In both cases, points conceded are encoded as horizontal position,
points scored are encoded as vertical position, and hemisphere
is encoded as shape.
(The only other difference is the scale; the transformation
from a hemisphere value to a shape).

Figure \ref{fig:textaes}\subref*{fig:textaes-2}
shows yet another version of the scatter plot that encodes the qualitative
hemisphere data values as both the colour of the text characters
as well as the shape of the text characters. Again, we see that,
when we use a character as the data symbol,
we can still use standard visual features, like colour, to encode
data values.

\begin{CodeChunk}


\begin{flushleft}\includegraphics[width=0.4\linewidth]{qual-char-mapping} \includegraphics[width=0.4\linewidth]{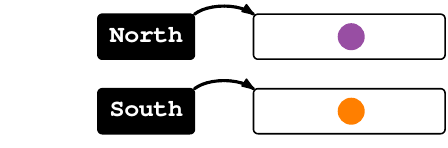} \end{flushleft}

\end{CodeChunk}

\begin{CodeChunk}
\begin{figure}

{\centering \subfloat[The data symbols are data points and the hemisphere of the teams is encoded as the shape of the points. We can decode the hemisphere for each team based on the shape of the data points.\label{fig:textaes-1}]{\includegraphics[width=0.49\linewidth]{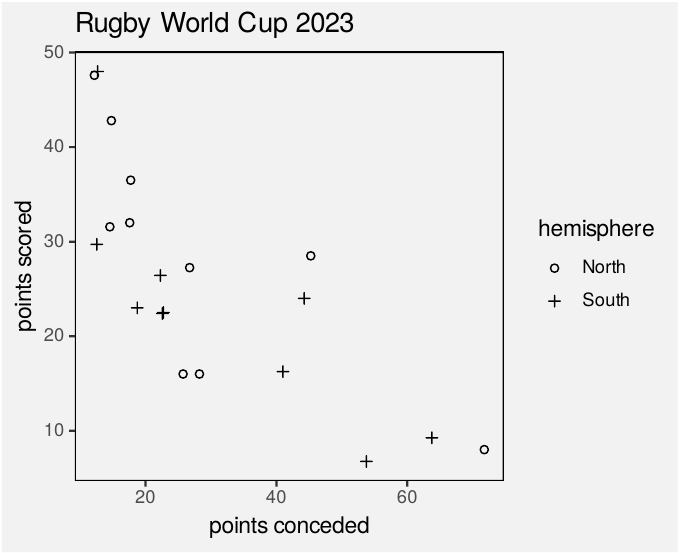} }\subfloat[The data symbols are text characters the hemisphere of the teams is encoded as the colour of the characters.  We can decode the hemisphere of each team based on the colour of the characters as well as the shape of the characters.\label{fig:textaes-2}]{\includegraphics[width=0.49\linewidth]{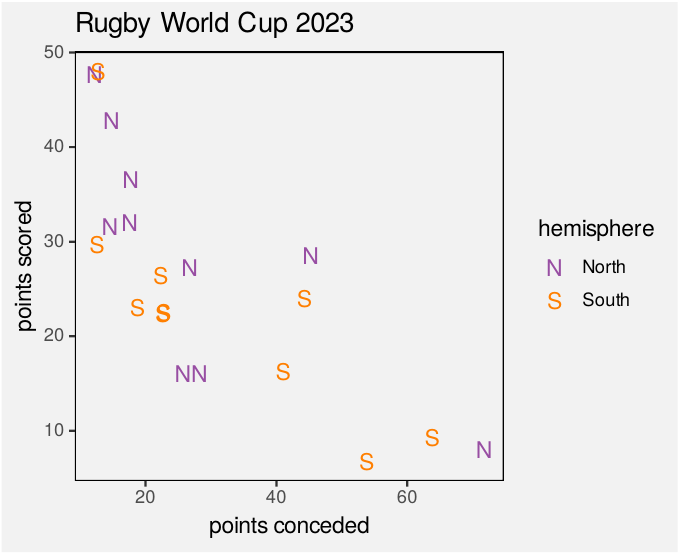} }

}

\caption[Scatter plots of the number of points scored and the number of points conceded (both are per-game averages) for teams in the 2023 Rugby World Cup]{Scatter plots of the number of points scored and the number of points conceded (both are per-game averages) for teams in the 2023 Rugby World Cup.}\label{fig:textaes}
\end{figure}
\end{CodeChunk}

In summary, if we use a text character as a data symbol,
we can encode data values as visual features of the text---position,
colour, and shape---just as we can for any other data symbol.

\hypertarget{text-specific-visual-features}{%
\subsection{Text-specific visual features}\label{text-specific-visual-features}}

The visual appearance of text can be controlled by many different parameters.
A character, like ``N'', is really just an abstract shape, with
many visual variations possible,
based on the font family, font face, font weight,
font style, and so on \citep{BRATH201659}.
This leads to one difference when we use text as a data symbol:
there are text-specific visual features that we can use to encode data
values.
For example,
Figure \ref{fig:textaes2}
shows a version of the scatter plot
with text characters as the data symbols and hemisphere encoded
as the font face of the character (as well as the shape of the character).
The ``N'' character for northern-hemisphere teams is drawn bold, while
the ``S'' character is drawn in a plain face.

\begin{CodeChunk}


\begin{flushleft}\includegraphics[width=0.4\linewidth]{qual-char-mapping} \includegraphics[width=0.4\linewidth]{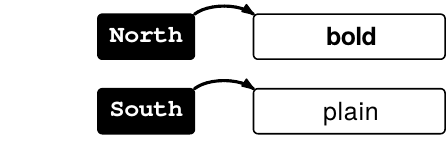} \end{flushleft}

\end{CodeChunk}

\begin{CodeChunk}
\begin{figure}

{\centering \includegraphics[width=0.49\linewidth]{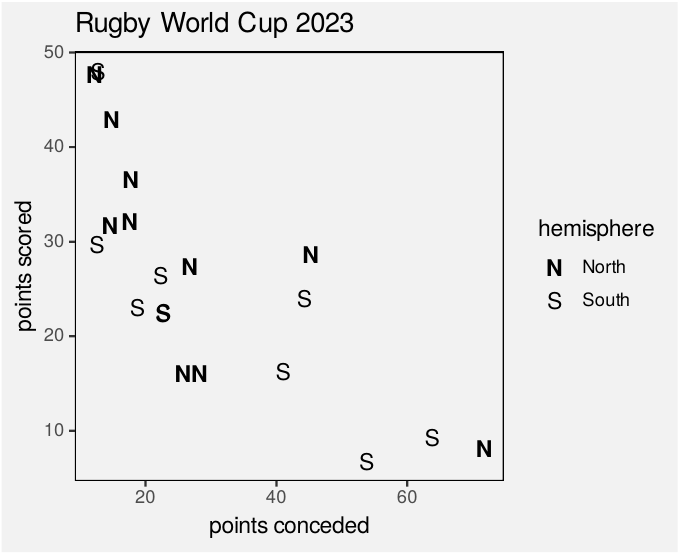} \includegraphics[width=0.49\linewidth]{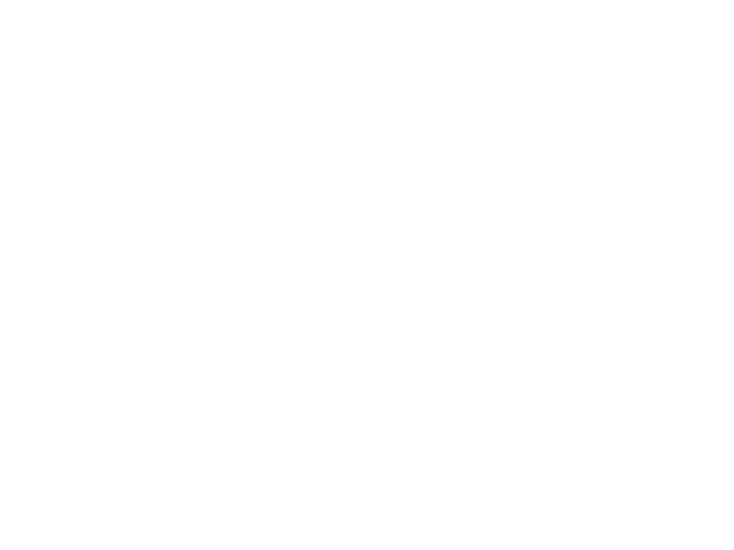} 

}

\caption[A scatter plot of the number of points scored and the number of points conceded (both are per-game averages) for teams in the 2023 Rugby World Cup]{A scatter plot of the number of points scored and the number of points conceded (both are per-game averages) for teams in the 2023 Rugby World Cup.  The data symbols are text characters with different font weights. We can decode the hemisphere of each team based on the font weight of the characters as well as the shape of the characters.}\label{fig:textaes2}
\end{figure}
\end{CodeChunk}

In summary, using text as a data symbol is, in many ways, just like
using any other data symbol.
Just as we can encode a data value as the colour of a bar,
we can encode a data value as the colour of text.
Similarly, we can encode data values as the position, shape, and size of
text.
The main difference
is that, when we use text as the data symbol,
there are some text-specific visual features, like font weight,
that we can also make use of.

\hypertarget{sec:words}{%
\section{Words as data symbols}\label{sec:words}}

It is also possible to encode data values as whole words.
For example, Figure \ref{fig:word}\subref*{fig:word-1}
shows a version of the scatter plot
with words as the data symbols.
There are several important points about this version of the scatter plot.

First of all, the data values are again encoded as the \emph{shape} of the text.
Different data values are encoded as different text shapes.
Instead of just a single character, a word is a collection
of characters, in effect a collection of shapes.
Put another way, a word is a more complex shape than a single
character.

Second of all, there is still a scale involved.
Data values are being transformed into words.
A word is a collection of characters and
the different hemisphere values are being encoded as
character versions of the data values.

\begin{CodeChunk}


\begin{flushleft}\includegraphics[width=0.4\linewidth]{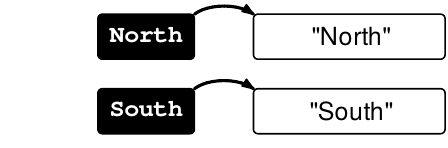} \end{flushleft}

\end{CodeChunk}

However, this is just one possible choice for the collection of characters.
The transformation does not have to use these particular
collections of characters.
To make this point clear, Figure \ref{fig:word}\subref*{fig:word-2}
shows another version
of the scatter plot in which the hemisphere values are encoded as
different words, but using a different scale
(a different transformation from data values to collections of characters).
The point is that the choice of text shapes is a deliberate choice,
not a forced choice.

\begin{CodeChunk}


\begin{flushleft}\includegraphics[width=0.4\linewidth]{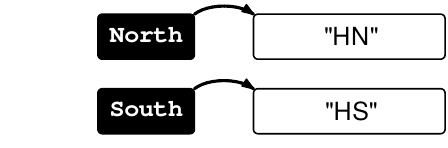} \end{flushleft}

\end{CodeChunk}

The third point is that the choice of ``North'' and ``South'' as the
text shapes, the choice of those particular collections of characters,
is a very good choice because of the
extremely powerful \emph{decoding}
that this provides.
On one hand, the encoding of different data values as different words
just produces
data symbols with different shapes (collections of different characters
of various shapes).
But for some shapes, for certain collections of characters,
human viewers do not just perceive a set of different shapes,
we also process those shapes to obtain meaning (semantic content).
For example, in this case
we decode the data value \texttt{North} from the
word ``North''.

\begin{CodeChunk}


\begin{flushleft}\includegraphics[width=0.4\linewidth]{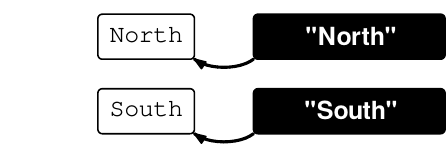} \end{flushleft}

\end{CodeChunk}

We can see this in Figure \ref{fig:word}\subref*{fig:word-1} because
there is no need for a legend. There is no need to explain what
the data symbol ``North'' means.
By contrast, in Figure \ref{fig:word}\subref*{fig:word-2} we still
need a legend to explain the encoding because ``HN'' is a collection
of characters that has no semantic content.

This is the most important feature of encoding data values as the
shapes of words:
The decoding is effective because of \emph{learned} associations
between particular collections of character shapes and their
semantic content. The viewer is able to read and
comprehend data symbols that
are words.

\begin{CodeChunk}
\begin{figure}

{\centering \subfloat[The data symbols are words. We can decode the hemisphere from the shape of the words. There is no need for a legend.\label{fig:word-1}]{\includegraphics[width=0.49\linewidth]{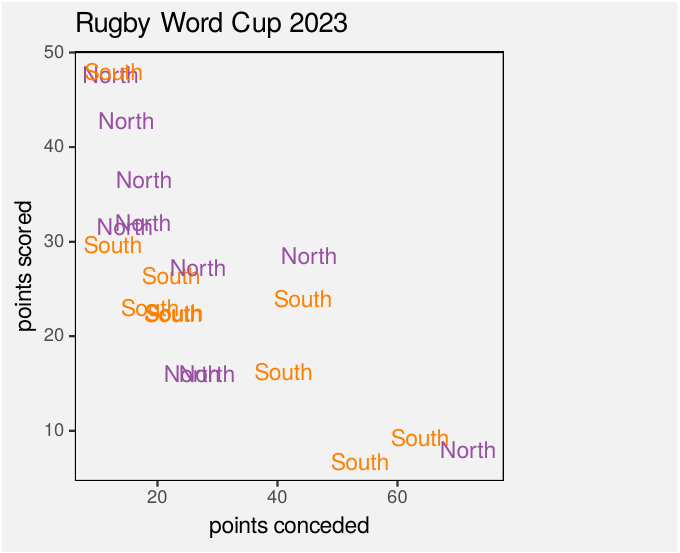} }\subfloat[The data symbols consist of multiple characters, but they are not recognisable as words.  We still need a legend to decode the shapes of the characters to the associated hemispheres.\label{fig:word-2}]{\includegraphics[width=0.49\linewidth]{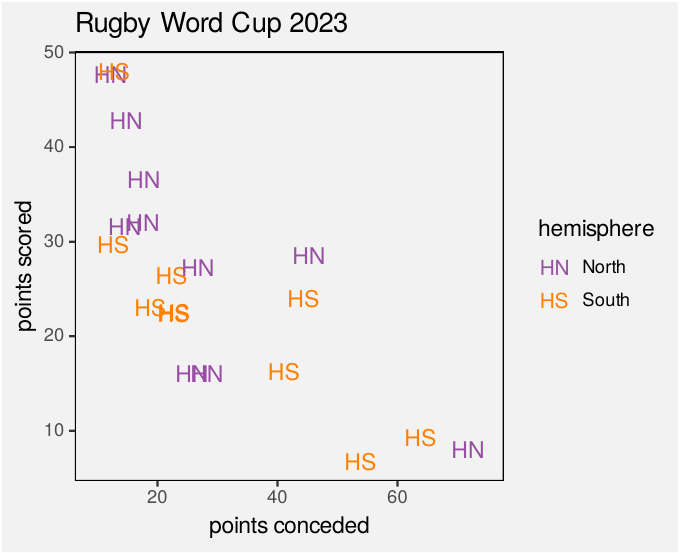} }

}

\caption[Scatter plots of the number of points scored and the number of points conceded (both are per-game averages) for teams in the 2023 Rugby World Cup]{Scatter plots of the number of points scored and the number of points conceded (both are per-game averages) for teams in the 2023 Rugby World Cup.}\label{fig:word}
\end{figure}
\end{CodeChunk}

\hypertarget{words-as-axis-labels}{%
\subsection{Words as axis labels}\label{words-as-axis-labels}}

Figure \ref{fig:wordAxis}\subref*{fig:wordAxis-1}
shows a more familiar use of text
to represent data values. This figure shows a bar plot
of the total number of male and female offenders in 2024
(an aggregation of a subset of the data in Figure \ref{fig:drama}).
For our current purposes,
the important text data symbols in this plot are the labels on the x-axis.

The data values in Figure \ref{fig:wordAxis}\subref*{fig:wordAxis-1}
are two quantitative counts and two
qualitative categories, \texttt{Male} and \texttt{Female}.
The quantitative encoding is quite straightforward:
the counts are encoded as the heights of the bars.
However, there is a lot more going on with the qualitative encoding.

\begin{CodeChunk}
\begin{figure}

{\centering \subfloat[Each bar has a word label.  We can decode the sex of the offender from the labels.\label{fig:wordAxis-1}]{\includegraphics[width=0.49\linewidth]{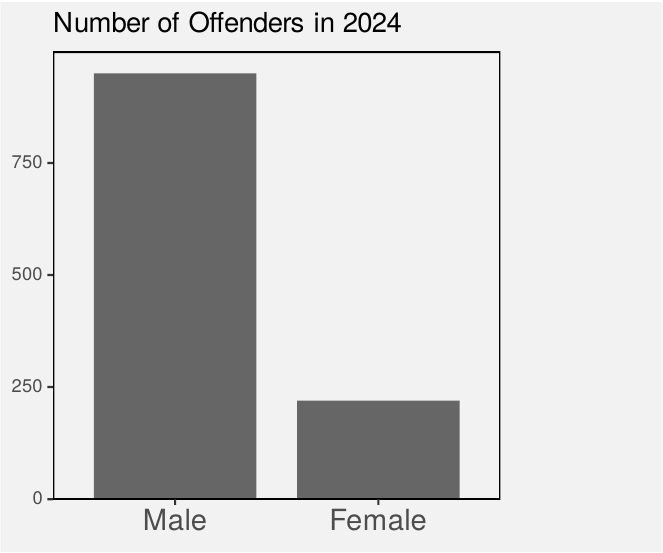} }\subfloat[The bars have no labels.  The positions of the bars indicate different groups, but not the identity of the groups.\label{fig:wordAxis-2}]{\includegraphics[width=0.49\linewidth]{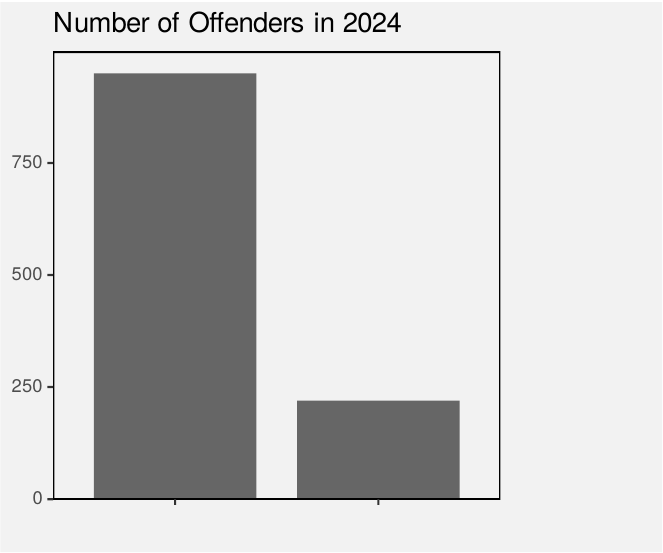} }

}

\caption[Bar plots of the number of male youth offenders and the number of female youth offenders in New Zealand in 2024]{Bar plots of the number of male youth offenders and the number of female youth offenders in New Zealand in 2024.}\label{fig:wordAxis}
\end{figure}
\end{CodeChunk}

The qualitative data values are encoded as the visual features of
two data symbols: bars and text words.
For the bars, the different values, \texttt{Male} and \texttt{Female}, are encoded as
different horizontal positions of the bars: the bar for males is to the left
of the bar for females.

\begin{CodeChunk}


\begin{flushleft}\includegraphics[width=0.4\linewidth]{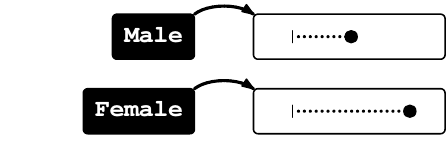} \end{flushleft}

\end{CodeChunk}

For the word data symbols,
the different values, \texttt{Male} and \texttt{Female}, are encoded as
the horizontal positions
of the text \emph{and} as the shapes of the text.
For example, the word ``Male'' is used to represent the data value \texttt{Male}
and that word is positioned on the left, below the left bar.

\begin{CodeChunk}


\begin{flushleft}\includegraphics[width=0.4\linewidth]{qual-pos-mapping} \includegraphics[width=0.4\linewidth]{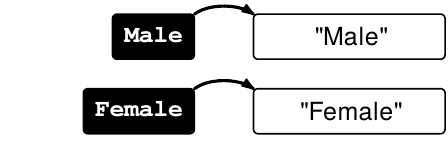} \end{flushleft}

\end{CodeChunk}

Now consider the \emph{decodings} from visual features back to
(the qualitative) data values in
Figure \ref{fig:wordAxis}\subref*{fig:wordAxis-1}.
The bars are at different horizontal positions, which indicates that there are
two different data values (two groups).
However, this decoding does not tell us what those groups are.
To make this clear,
Figure \ref{fig:wordAxis}\subref*{fig:wordAxis-2}
shows a version of the bar plot \emph{without} the text word axis labels.
All we can tell from
Figure \ref{fig:wordAxis}\subref*{fig:wordAxis-2}
is that the bars are at different positions, which suggests
that there are two different groups, but the decoding just from
the positions of the bars does
not tell us that those groups are
\texttt{Male} and \texttt{Female}.

\begin{CodeChunk}


\begin{flushleft}\includegraphics[width=0.4\linewidth]{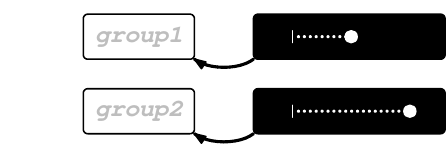} \end{flushleft}

\end{CodeChunk}

Similarly, the text labels are at two different positions,
which indicates that there are
two different groups, without saying what the groups are.
However, the most important decoding is the learned decoding
from the text shapes to the semantic concepts of a \texttt{Male} group
and a \texttt{Female} group.

\begin{CodeChunk}


\begin{flushleft}\includegraphics[width=0.4\linewidth]{qual-pos-inv-mapping} \includegraphics[width=0.4\linewidth]{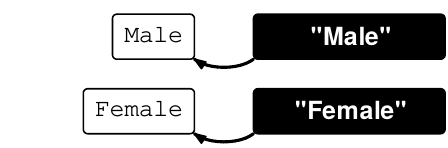} \end{flushleft}

\end{CodeChunk}

The most important encoding of \texttt{Male} and \texttt{Female} is as the shape
of the text words in the x-axis labels. This allows
decoding back to \texttt{Male} and \texttt{Female}.
The proximity of those text labels to the bars---the
consistent encoding of \texttt{Male} and \texttt{Female} to the position of both
the bars and the text words---forms visual groupings \citep{Todorovic:2008}
and that allows us to
connect the different bars with the \texttt{Male} and \texttt{Female} groups that
we are able to decode from the x-axis labels .

\newpage

\hypertarget{words-as-legend-labels}{%
\subsection{Words as legend labels}\label{words-as-legend-labels}}

Figure \ref{fig:direct}\subref*{fig:direct-1} shows a line
plot of the number of \texttt{Male} and \texttt{Female} offenders over three years
(an aggregation of a subset of the data from Figure \ref{fig:drama}).
The data values that we will focus on are the two groups
\texttt{Male} and \texttt{Female}.
These values are encoded as the colour of the lines and points,
but that only allows us to decode that there are two different groups.

\begin{CodeChunk}


\begin{flushleft}\includegraphics[width=0.4\linewidth]{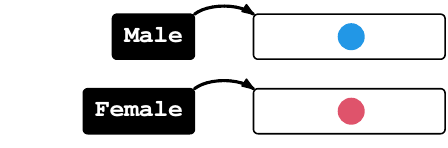} \includegraphics[width=0.4\linewidth]{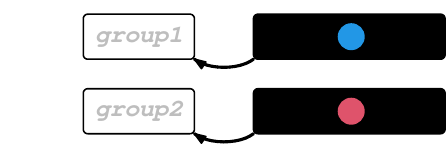} \end{flushleft}

\end{CodeChunk}

However, in addition, the data values \texttt{Male} and \texttt{Female}
are encoded as visual features of lines (and points) and text
words in a legend.
In this legend, the data values are encoded as the colour of the
lines and points, just as they are in the main plot,
plus they are encoded as the the shape of the words, ``Male'' and ``Female''.
The decoding from the coloured lines (and points)
only allows us to identify different groups, as in the main plot,
but the words allow us to decode that the groups are
\texttt{Male} and \texttt{Female}.

\begin{CodeChunk}


\begin{flushleft}\includegraphics[width=0.4\linewidth]{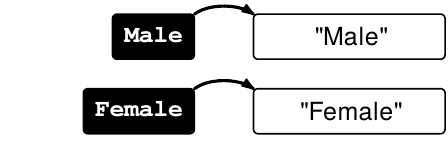} \includegraphics[width=0.4\linewidth]{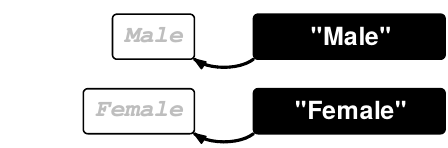} \end{flushleft}

\end{CodeChunk}

The proximity of the lines (and points) and text in the legend
forms a visual grouping that allows us to associate the data values
that we decode from the text with the coloured lines.
The similarity of the lines (and points) in the legend and in the main
plot forms another visual grouping that allows us to transfer that
association to the lines in the main plot \citep{Todorovic:2008}.

\begin{CodeChunk}
\begin{figure}

{\centering \subfloat[The sex of the offender is encoded as the colour of the lines and there is a legend to explain the encoding.  We can decode the sex of the offender from the labels in the legend and use proximity to group the labels with the lines in the legend and similarity to group the lines in the legend with the lines in the plot.\label{fig:direct-1}]{\includegraphics[width=0.49\linewidth]{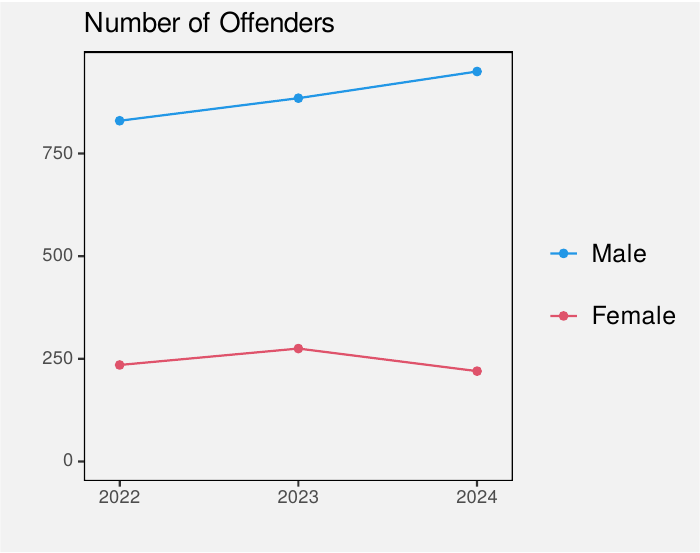} }\subfloat[The sex of the offender is encoded as the colour of the lines and as the colour of text data symbols (direct labels).  We can decode the sex of the offender from the labels in the plot and use proximity and similarity to group the labels with the lines in the plot.\label{fig:direct-2}]{\includegraphics[width=0.49\linewidth]{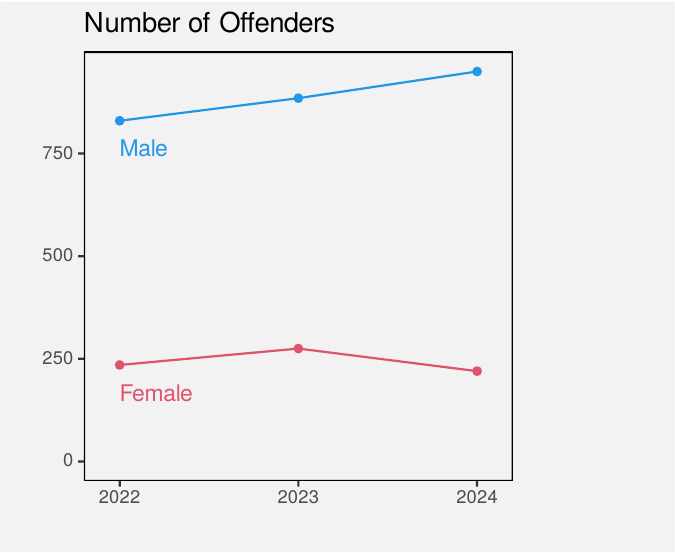} }

}

\caption[Line plots of the number of male youth offenders and the number of female youth offenders in New Zealand from 2022 to 2024]{Line plots of the number of male youth offenders and the number of female youth offenders in New Zealand from 2022 to 2024.}\label{fig:direct}
\end{figure}
\end{CodeChunk}

This use of words in the legend of
Figure \ref{fig:direct}\subref*{fig:direct-1}
is similar to what we saw with
words in the x-axis in
Figure \ref{fig:wordAxis}\subref*{fig:wordAxis-1}, but
with more complicated visual groupings required to decode
the data values from the data symbols.

Figure \ref{fig:direct}\subref*{fig:direct-2}
shows a simpler alternative: direct labelling.
In this version of the line plot, a text word is placed
next to the relevant line within the plot itself.
The encodings in
Figure \ref{fig:direct}\subref*{fig:direct-2}
are very similar to the encodings in
Figure \ref{fig:direct}\subref*{fig:direct-1},
but there are no extra lines (and points) in a separate legend
and the \texttt{Male} and \texttt{Female} data values are encoded as both
the shape and colour of the words within the plot.
The decoding of data values from the word shapes is
associated with the lines (and points) through
a single visual grouping that is very strong because
it is based on both the proximity of the words and lines and
the repeated colour of the words and lines.

Figure \ref{fig:direct}\subref*{fig:direct-2} also provides a very nice
demonstration of the value of thinking of text as just another data symbol
that we can use to encode data values.
The direct labels are text data symbols with the sex of the offender
encoded as the position, shape, and colour of the text.

In summary, the main benefit of using text as a data symbol
for qualitative data is
the ability to \emph{decode} not just that there are different groups,
but to obtain different \emph{semantic content} about the groups
from the text.
This is essential to the effectiveness of qualitative axes, legends,
and direct labels in a
data visualisation.

\hypertarget{sec:numbers}{%
\section{Numbers as data symbols}\label{sec:numbers}}

Numbers are another important type of text data symbol.
The text in this case is a collection of digits (plus possibly
decimal points and perhaps a comma group separator).
For example,
Figure \ref{fig:number}\subref*{fig:number-1}
shows a variation of the bar plot of the number
of male versus female offenders, with a white text label at the top of
each bar to
show the exact count for each bar.

\begin{CodeChunk}
\begin{figure}

{\centering \subfloat[The counts are encoded as the shape of the white numeric labels. We can decode the counts from the white labels.\label{fig:number-1}]{\includegraphics[width=0.49\linewidth]{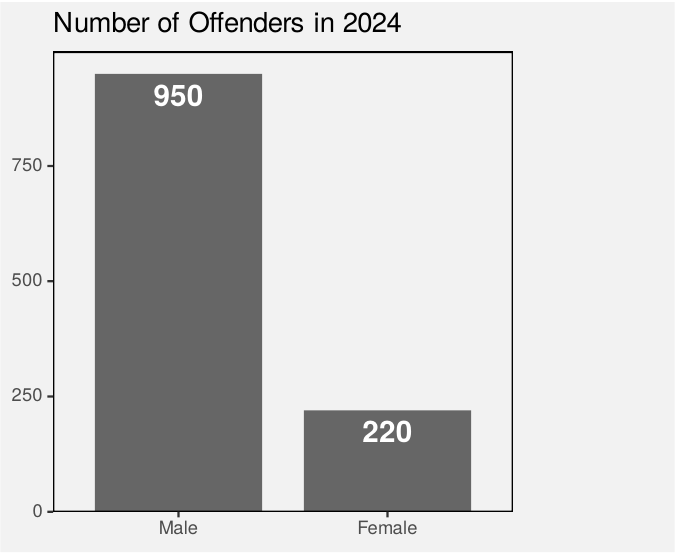} }\subfloat[The counts are encoded only as the heights of the bars. We can decode the relative heights of the bars, but not the absolute counts represented by the bars.\label{fig:number-2}]{\includegraphics[width=0.49\linewidth]{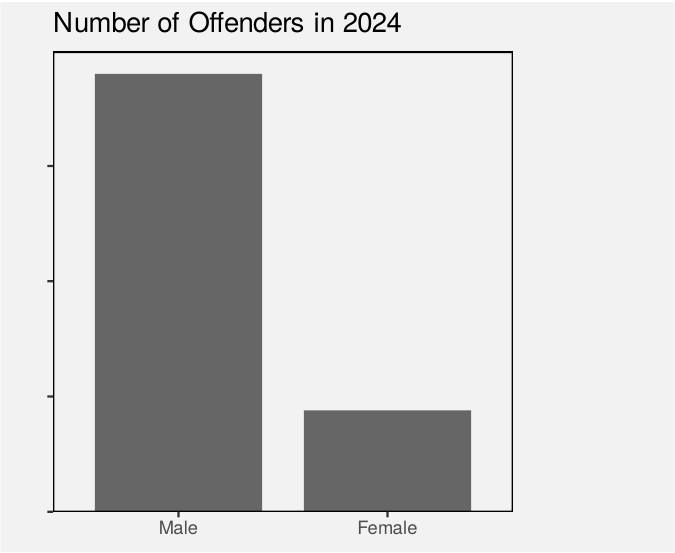} }

}

\caption[Bar plots of the number of male youth offenders and the number of female youth offenders in New Zealand in 2024]{Bar plots of the number of male youth offenders and the number of female youth offenders in New Zealand in 2024.}\label{fig:number}
\end{figure}
\end{CodeChunk}

There are two quantitative data values in
Figure \ref{fig:number}\subref*{fig:number-1}:
\texttt{950}, which is the count for \texttt{Male} offenders, and
\texttt{220}, the count for \texttt{Female} offenders.
If we just consider the white text data symbols for now,
the counts are encoded as
the vertical position and the shape of the white text numbers.
Again, there are scales involved to transform the data values to
visual features of the text, both to position the text
and to determine the shape of the text.
As with encoding data values as words, or collections of characters,
there are plenty of options for encoding data values as numbers,
or collections of digits.
For example, in this case, the data value
\texttt{950} is encoded as the text shape ``950'', but other possibilities are
``950.0'' and ``9.5e2''.
The point is that, as with text words, we make a deliberate choice
about the collections of digits that we use to encode numeric data values.

\begin{CodeChunk}


\begin{flushleft}\includegraphics[width=0.4\linewidth]{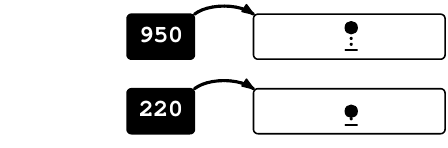} \includegraphics[width=0.4\linewidth]{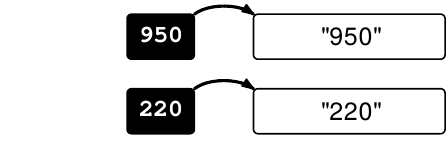} \end{flushleft}

\end{CodeChunk}

We can decode from the positions and shapes of the white
text numbers that there are
two different amounts, but the positions do not by themselves
tell us the exact amounts.
However, we can decode the exact counts from
the learned semantic content of the white text numbers.

\begin{CodeChunk}


\begin{flushleft}\includegraphics[width=0.4\linewidth]{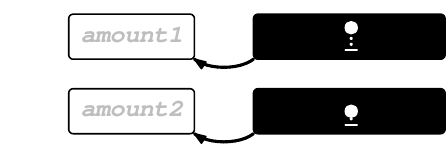} \includegraphics[width=0.4\linewidth]{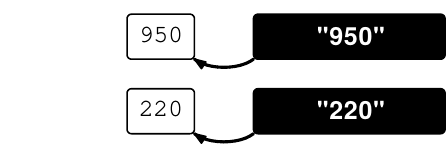} \end{flushleft}

\end{CodeChunk}

\hypertarget{numbers-as-axis-labels}{%
\subsection{Numbers as axis labels}\label{numbers-as-axis-labels}}

In
Figure \ref{fig:number}\subref*{fig:number-1},
the count data values are also encoded as the heights of bar data symbols.

\begin{CodeChunk}


\begin{flushleft}\includegraphics[width=0.4\linewidth]{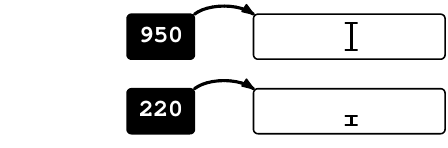} \end{flushleft}

\end{CodeChunk}

We can decode from the heights of the bars that the counts are
different \emph{and} that the smaller bar is approximately one-quarter
of the larger bar. This reflects the fact that height (or length)
is a quantitative visual feature, so we are able to decode the relative size
of differences, rather than just the fact that the heights are different,
which was all we could tell from different colours.
However, we are unable to decode the exact data values from
the heights of the bars alone. This is demonstrated in Figure
\ref{fig:number}\subref*{fig:number-2},
which has removed the y-axis tick labels to show that
we can then only decode relative size from the two bars.

\begin{CodeChunk}


\begin{flushleft}\includegraphics[width=0.4\linewidth]{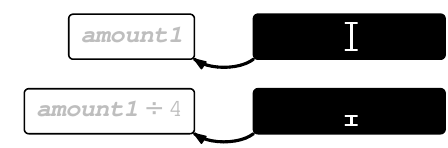} \end{flushleft}

\end{CodeChunk}

Figure \ref{fig:number}\subref*{fig:number-1}
contains other number data symbols besides the white text numbers;
there are also
y-axis tick labels.
The position and shape of these data symbols
allow us to decode the data values \texttt{0}, \texttt{250}, \texttt{500}, and \texttt{750} and
the position of the bar heights relative to those tick marks
allows us to decode the heights of the bars, not just as different amounts,
but as the approximate data value counts.

In summary, text data symbols that are numbers are as important as
text data symbols that are words,
and for the same reason: if we encode data values as
text numbers then we can use the semantic value
of the numbers to decode back to the original data values.
In other words, we are able to read and comprehend the text numbers.
This is essential to the effectiveness of numeric
labels and quantitative axes.

\hypertarget{sec:effectiveness}{%
\section{The effectiveness of text data symbols}\label{sec:effectiveness}}

The previous sections have shown that encoding data values
as the shape of text data symbols can be
very effective because it allows us to decode from a visual representation
back to data values. This helps to explain why data visualisations
contain text elements, particularly on axes and legends.

We have also hinted at the fact that decoding data values from the visual
features of other data symbols,
such as the positions of data points or the lengths of bars, is
limited compared to decoding data values from words or numbers.
For example, the horizontal positions of the bars in
Figure \ref{fig:wordAxis}\subref*{fig:wordAxis-1} allow us
to decode that there are two different groups, but not what the
groups are, whereas the shapes of the word labels below the bars in
Figure \ref{fig:wordAxis}\subref*{fig:wordAxis-1} allow us
to decode not only that there are two groups, but also that the
groups are \texttt{Male} and \texttt{Female}.

We have moved from questioning why text is present in a data visualisation
to wondering why we do not see even more text in
data visualisations? Why do we see so many geometric data symbols being
used in data visualisations?

The short answer is that decoding single data values from
data symbols is not the only task that we perform when
we view a data visualisation.
We need to consider other metrics for assessing the effectiveness
of different encodings and different data symbols in order to get
a more complete picture of where text performs well and where text
performs poorly.

A lot is already known about the effectiveness of
encoding data values as the visual features of simple
geometric symbols \cite[e.g.,][]{munzner2014visualization, 
ware2020information} and there are well-established
approaches to measuring the effectiveness of different encodings
\citep{annurev:/content/journals/10.1146/annurev-statistics-031219-041252}.
In this section, we will explore different ways to evaluate the
effectiveness of text data symbols, which will help to explain
why text is only used for specific roles within a data visualisation.

\hypertarget{expressiveness}{%
\subsection{Expressiveness}\label{expressiveness}}

In Figure \ref{fig:drama}, the counts of offenders are encoded
as the lengths of bars and the sex of the offenders is encoded
as the colours of the bars (pink versus blue).
One reason for those choices is the fact that lengths are appropriate
for representing quantitative values because our perception
of length is quantitative; we can perceive lengths as larger or smaller
and we can perceive the size of differences between lengths
and even the ratio of
differences between lengths.
Conversely, colour, or more accurately, hue, is appropriate
for representing qualitative values because our perception of
hue is qualitative; we can perceive different hues as distinct,
but without any implication of an ordering of hues.

Certain visual features
are appropriate for encoding data values of a certain type
because we are capable of decoding the important properties of the
encoded data values
from the visual features \citep[no more and no less;][]{ZHANG199659}.
This is sometimes referred to as the \emph{expressiveness} of
the visual feature \citep{munzner2014visualization}.

How well do text data symbols perform in terms of expressiveness?
First of all, expressiveness is a property of different visual features
rather than different data symbols, so the question really is:
how well do visual features of text data symbols perform in terms
of expressiveness?
As noted previously, when we employ text as a data symbol,
we can encode data values using many of the
visual features, such as colour, that we use with other data symbols
so in that sense text data symbols are as expressive as any other data
symbol for representing different types of data.
We can represent qualitative data values using qualitative
visual features, for example, text with different colours,
and we can represent quantitative data values using quantitative
visual features, for example, text at different positions.

However, we have also identified that
the shape of text data symbols is a very significant
visual feature because it decodes data values excellently thanks
to learned decoding of semantic content.
In this sense, text is different to other data symbols.
When we employ a data point as a data symbol,
encoding data values as the shape of the data symbol
is only effective for qualitative data,
like in Figure \ref{fig:textaes}\subref*{fig:textaes-2}.
We perceive different point shapes as different groups, but with
no inherent order.
If we encode data values as the shape of a single text character,
like in Figure \ref{fig:rwc}\subref*{fig:rwc-2},
the result is similar.
We only perceive different character shapes as different groups.
However, if we encode data values as the shape of words or numbers,
like in Figure \ref{fig:word}\subref*{fig:word-1},
then we perceive the semantic content of the words or numbers.
We decode the text shapes to specific data values.

Because the shape of text data symbols encompasses both words
and numbers, this means that text is appropriate for
representing both quantitative and qualitative data values.
We can see this in Figure \ref{fig:wordAxis}\subref*{fig:wordAxis-1}
with the use of text labels on both the qualitative x-axis
and the quantitative y-axis.

In addition, there are text-specific visual features, such as font
weight and font slant. Some of these are quantitative visual features,
such as font weight---a bolder font appears ``larger'' than a less bold
font---and some are qualitative visual features, such as font family
\citep{BRATH201659}.
In effect these text-specific visual features mean that text data
symbols provide
additional expressiveness compared to other data symbols.

Another situation where text data symbols are particularly appropriate
is when the data set consists of text, as in natural language
processing \citep{brath2020visualizing}.

In summary, text data symbols compare favourably to any other
sort of data symbol in terms of expressiveness.
We can encode both quantitative and qualitative
data values using both generic visual features, like colour, and text-specific
visual features, like font weight, plus we can
encode any sort of data value using text shape.

\hypertarget{capacity}{%
\subsection{Capacity}\label{capacity}}

Figure \ref{fig:rwcCountry}\subref*{fig:rwcCountry-1}
shows a version of the scatter plot of
points scored versus points conceded for teams in the 2023 Rugby World Cup.
The data symbol is a data point and the country name is encoded as
the colour of the data point.

\begin{CodeChunk}
\begin{figure}

{\centering \subfloat[The country is encoded as the colour of the data points.  It is difficult to differentiate between the colours.\label{fig:rwcCountry-1}]{\includegraphics[width=0.49\linewidth]{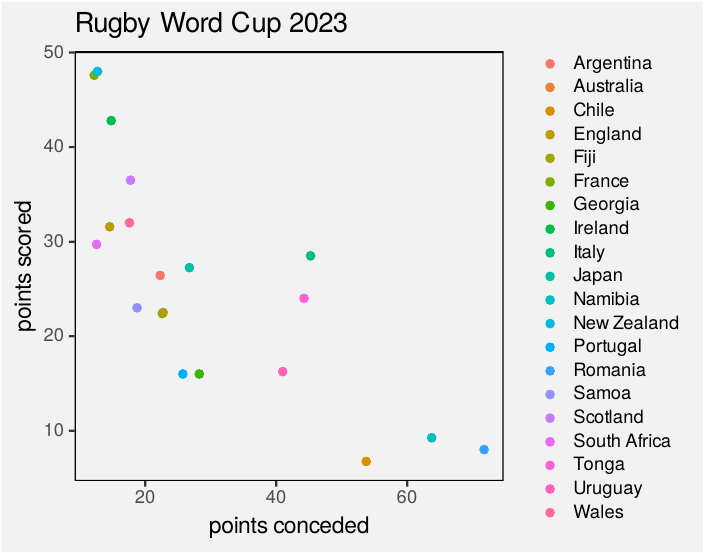} }\subfloat[The country name is encoded as the shape of the text data symbols.  It is very easy to differentiate between different country names, though overlapping makes the task significantly more difficult.\label{fig:rwcCountry-2}]{\includegraphics[width=0.49\linewidth]{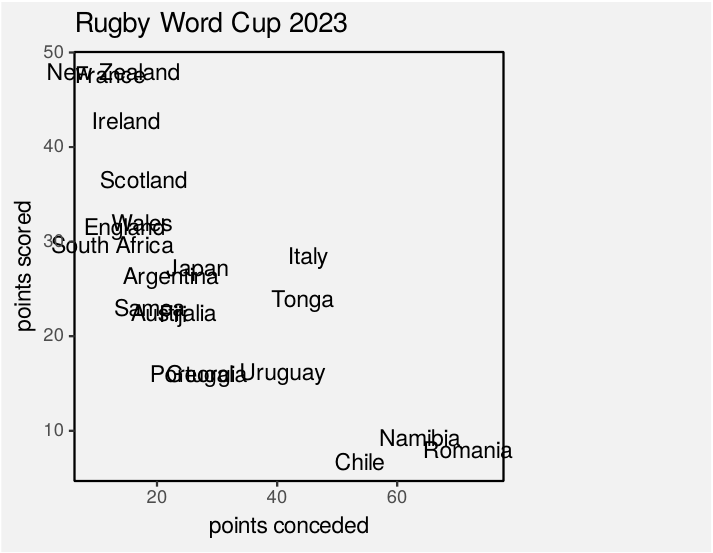} }

}

\caption[Scatter plots of the number of times that a team against through the opposition defence and the number of for that a team scores (both are per-game averages) for teams in the 2023 Rugby World Cup]{Scatter plots of the number of times that a team against through the opposition defence and the number of for that a team scores (both are per-game averages) for teams in the 2023 Rugby World Cup.}\label{fig:rwcCountry}
\end{figure}
\end{CodeChunk}

Figure \ref{fig:rwcCountry}\subref*{fig:rwcCountry-1}
shows a limitation of colour as a visual feature:
there is a limit to how many different colours we can use at once
\citep{ware2020information,munzner2014visualization}.
The decoding from colours to country names is difficult because
it is difficult to differentiate between some of the colours,
particularly when they are scattered in space like the data symbols,
but even when they are carefully arranged like in the legend.
For example, can you differentiate between
the points that correspond to Argentina,
Australia, Uruguay, and Wales?

We can compare visual features in terms of their \emph{capacity}---how
many different levels of the visual feature can be
differentiated---particularly
when encoding different categories for qualitative data.
For example, position has a much higher capacity than colour,
as we can see in the
legend for Figure \ref{fig:rwcCountry}\subref*{fig:rwcCountry-1};
we have no difficulty
differentiating between the different vertical positions of the 20
data points and country names in the legend.

For visual features like colour and position,
text data symbols have the same capacity as any other data symbol.
We can represent a small number of different categories
using text with different colours, like in
Figure \ref{fig:word},
and we can represent a larger number of different categories
using text at different positions, like in the legend for
Figure \ref{fig:rwcCountry}\subref*{fig:rwcCountry-1}.

When we use data points as data symbols
(e.g., Figure \ref{fig:textaes}\subref*{fig:textaes-1}),
the capacity of shape as a visual feature is limited,
especially when data symbols overlap.
It rapidly becomes difficult to differentiate between symbol shapes
\citep{cleveland1994elements}.
Figure \ref{fig:rwcCountry}\subref*{fig:rwcCountry-2}
shows that the shape of text data symbols also suffers from
difficulties with overlap.
However, both
Figure \ref{fig:rwcCountry}\subref*{fig:rwcCountry-1} and
Figure \ref{fig:rwcCountry}\subref*{fig:rwcCountry-2}
show that the shape of text, when decoded to semantic content
has a very high capacity.
We are able to differentiate easily between all of the country names
in the legend of
Figure \ref{fig:rwcCountry}\subref*{fig:rwcCountry-1}
and between the non-overlapping data symbols in
Figure \ref{fig:rwcCountry}\subref*{fig:rwcCountry-2}.
In fact, it is difficult to think of a set of data values
that would exceed the limits of text shape.

When we use text as data symbols, there are also text-specific
visual features available. Although there is some evidence that
it is possible to effectively discriminate between, for example, different
levels of font weight, the number of different levels is no
larger than for colour \citep{10.1145/3528223.3530111}.

In summary, text data symbols are at least the equal of any other data symbol
in terms of capacity, when encoding data values as standard visual features
of text, like colour and position. Text data symbols offer some
additional flexibility with text-specific visual features and have
unmatched capacity when encoding data values as the shape of text.

\hypertarget{accuracy}{%
\subsection{Accuracy}\label{accuracy}}

Figure \ref{fig:area} presents the same data values as
the bar plot in
Figure \ref{fig:wordAxis}\subref*{fig:wordAxis-1},
but with different encodings.
In both cases, the sex of the offender is encoded as horizontal position.
However, in Figure \ref{fig:wordAxis}\subref*{fig:wordAxis-1}
the counts of offenders are encoded
as the heights of bars, whereas in Figure \ref{fig:area}
the counts are encoded as the \emph{area} of circles.

It is more difficult to decode the data values from the areas
of the circles in Figure \ref{fig:area}
compared to decoding the
data values from the heights of the bars in
Figure \ref{fig:wordAxis}\subref*{fig:wordAxis-1}.
It is also harder to decode the \emph{relative} areas of the circles
compared to decoding the relative heights of the bars.
In fact, there is an established ranking of visual features that
places position and length above visual features like area and angle
and shape in terms of the accuracy of decoding
\citep{cleveland1984graphical,10.1145/1753326.1753357}.

\begin{CodeChunk}
\begin{figure}

{\centering \includegraphics[width=0.49\linewidth]{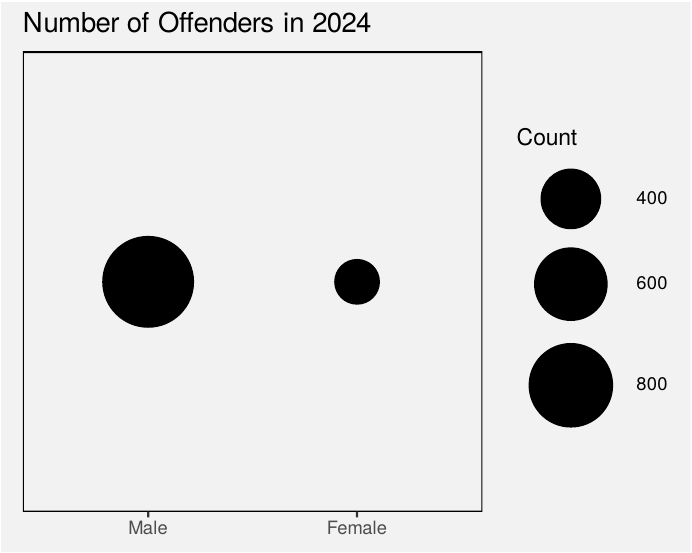} \includegraphics[width=0.49\linewidth]{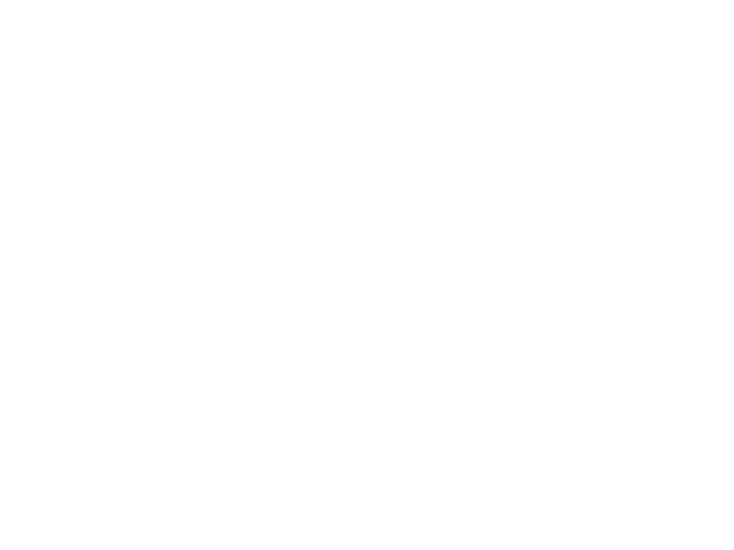} 

}

\caption[A bubble plot of the number of male youth offenders and the number of female youth offenders in New Zealand in 2024. The number of offenders is encoded as the size of each circle. We can decode the relative sizes of the circles, but not as well as we can decode the relative heights of bars, like in Figure \ref{fig:wordAxis}.]{A bubble plot of the number of male youth offenders and the number of female youth offenders in New Zealand in 2024. The number of offenders is encoded as the size of each circle. We can decode the relative sizes of the circles, but not as well as we can decode the relative heights of bars, like in Figure \ref{fig:wordAxis}.}\label{fig:area}
\end{figure}
\end{CodeChunk}

What is the accuracy of decoding when we encode data values
using text data symbols?
Again, we should talk about the accuracy of visual features
rather than data symbols and, in most cases, encoding a
data value as the visual feature of a text data symbol
will be just as accurate or innacurate as encoding the data value
as the same visual feature of a different data symbol.
However,
Figure \ref{fig:number}\subref*{fig:number-1}
shows that, if we encode data values as the \emph{shape}
of text data symbols, by adding a number label at the top of each bar,
then we can very accurately decode the original
data values. This decoding of text shape is far superior to the
decoding of shape for other data symbols and is even superior to decoding from
height or position. In that sense, the shape of text data symbols is an
extremely accurate visual feature.

However, the above assessment only focuses on
decoding from a single visual feature to a single data value.
While the ability to extract single data values has
value in some settings
\citep{ijgi9070415}
and arguably
underlies more complex decodings \citep{cleveland1984graphical},
comparisons between two or more values are both more common and more useful
when reading a data visualisation
\cite[p. 127]{10.1109/INFOVIS.2005.24,tufte2006beautiful}.

For example, the established ranking of visual features in terms of accuracy
is based on the perception of ratios of data values.
In Figure \ref{fig:wordAxis}\subref*{fig:wordAxis-1},
this corresponds to determining
the height of the bar for females relative to the height of the bar
for males.
In Figure \ref{fig:area},
this corresponds to decoding the area of the
female circle relative to the area of the male circle.
Making this comparison is more accurate when data values are
encoded as lengths or positions compared to when data values
are encoded as area or angle.

Is the comparison of two data values harder or easier
when the values are encoded as text shape?
It is possible to perform this comparison using the text
data symbols in Figure \ref{fig:number}\subref*{fig:number-1}
based on the text shape because learned decoding of numbers
means that we obtain exact numeric values 220 and 950.
We can mentally approximate 220 divided by 950
(somewhere between a quarter and a fifth) and we can
even calculate a more precise value for 220 divided by 950
(0.23). However,
mentally calculating 220 divided by 950 is slower and requires more
cognitive effort compared to decoding the relative heights of
the bars (at least to get an approximate proportion).

If we have more than two values to compare, the situation gets worse for text.
As the number of comparisons increases, comparing text numbers
gets even slower than comparing, for example, multiple bar lengths.
Text may still be an option in situations where geometric data symbols
are themselves not the ideal solution, for example, with
thematic maps \citep{ijgi9070415}, but geometric data symbols
will generally provide better support for multiple comparisons.

If we encode data values using text-specific visual features
(that are appropriate for quantitative data), such as
font weight, it is possible to decode data values, but not with
very great accuracy
\citep{10.1145/3528223.3530111}.

In summary, encoding data values as the shape of
text data symbols leads to extremely accurate decoding of
individual data values. However, the decoding is slow and just gets slower
when comparing
two or more values, which is much more common and much more useful.
We have encountered our first sign of weakness for text data symbols.

\hypertarget{learned}{%
\subsection{Learned decodings}\label{learned}}

Encoding data values as the shape of text is very effective because
learned decodings allow us to recover the original data values,
for example, the data value \texttt{Female} from the text ``Female''.
This property of text shape is fundamental to the effectiveness of text
in a data visualisation.
However, this property is not unique to the shape visual feature of
text data symbols.
For example, Figure \ref{fig:colLearned} shows a variation of the
bar plot of counts of offenders by the sex of the offender.
This plot has no x-axis label and no legend, but if forced to guess
which bar represents \texttt{Male},
there are learned associations that favour the left bar because
the colour blue is associated with male, while pink
is associated with female.

\begin{CodeChunk}
\begin{figure}

{\centering \includegraphics[width=0.49\linewidth]{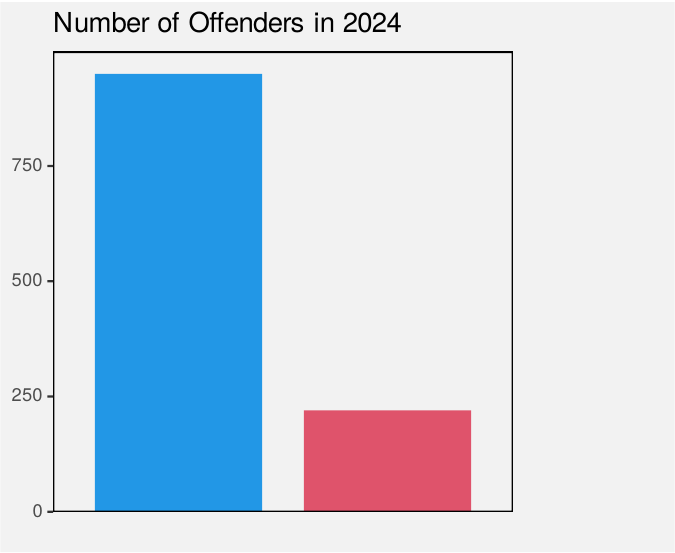} \includegraphics[width=0.49\linewidth]{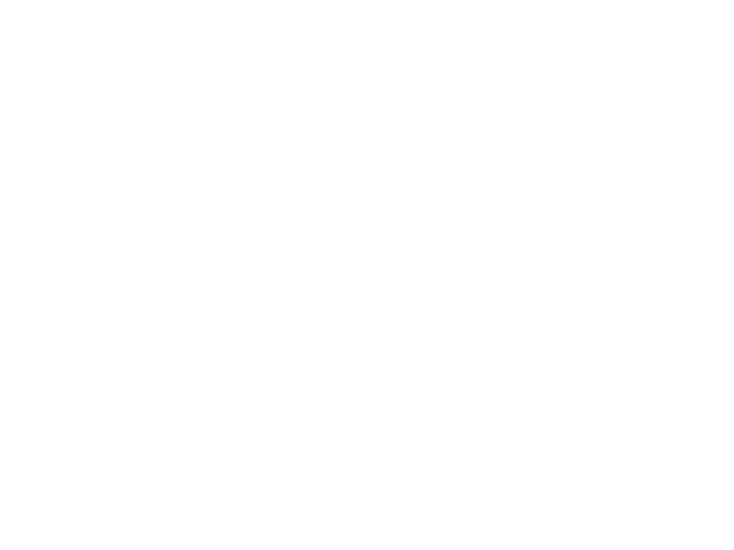} 

}

\caption[A bar plot of the number of male youth offenders and the number of female youth offenders in New Zealand in 2024]{A bar plot of the number of male youth offenders and the number of female youth offenders in New Zealand in 2024.  Although there are no x-axis labels, we can guess at the sex of the offender from the colour of the bars.}\label{fig:colLearned}
\end{figure}
\end{CodeChunk}

\begin{CodeChunk}


\begin{flushleft}\includegraphics[width=0.4\linewidth]{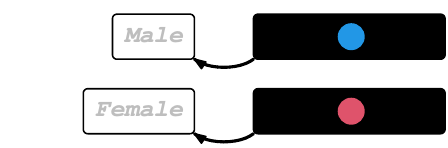} \end{flushleft}

\end{CodeChunk}

In other words, encoding a data value as the colour of a non-text data symbol
can be effective in a similar way to encoding a data value as the
shape of a text data symbol
because there are learned decodings of colour, just as there are
learned decodings of text shape.
However, this sort of learned decoding based on colour
tends not to be as effective
as a learned decoding based on
the shape of a text data symbol because the learned decodings based on
colour tend to
be more ambiguous. For example, there are other common semantic associations
with the colours pink and blue, like hot and cold.

\begin{CodeChunk}


\begin{flushleft}\includegraphics[width=0.4\linewidth]{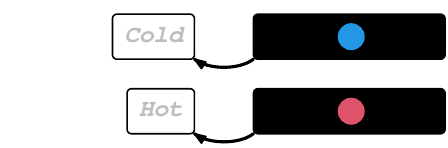} \end{flushleft}

\end{CodeChunk}

\begin{CodeChunk}
\begin{figure}

{\centering \subfloat[The x-axis labels that are symbols rather than words.  We can decode the sex of the offender from the symbols, although this decoding is less reliable than decoding from words.\label{fig:symbolLearned-1}]{\includegraphics[width=0.49\linewidth]{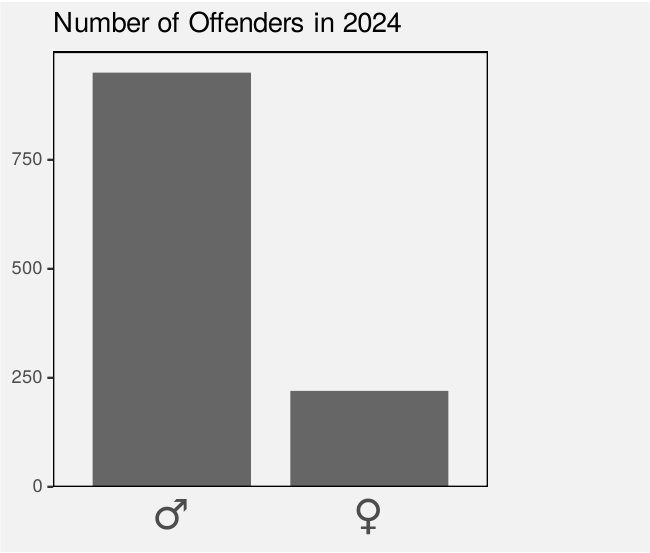} }\subfloat[The x-axis labels that are single characters rather than words. We can decode the sex of the offender from the single character, although this decoding is less reliable than decoding from whole words.\label{fig:symbolLearned-2}]{\includegraphics[width=0.49\linewidth]{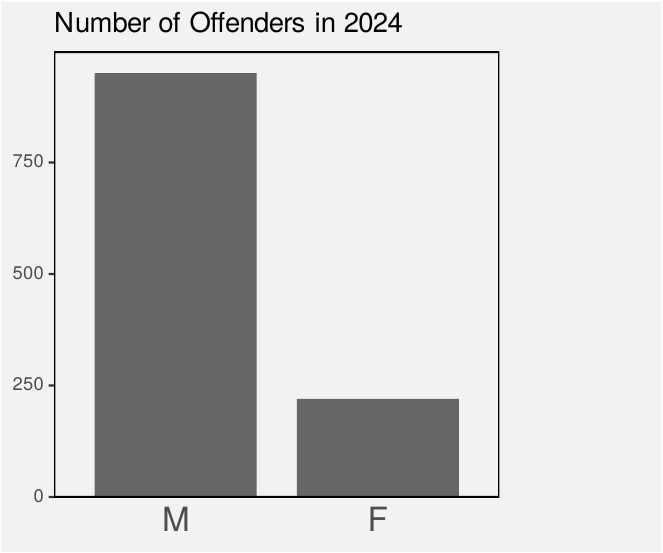} }

}

\caption[Bar plots of the number of male youth offenders and the number of female youth offenders in New Zealand in 2024]{Bar plots of the number of male youth offenders and the number of female youth offenders in New Zealand in 2024.}\label{fig:symbolLearned}
\end{figure}
\end{CodeChunk}

Figure \ref{fig:symbolLearned} shows two other examples
of encoding data values as the shape of data symbols, where the data
symbols are not words.
In Figure \ref{fig:symbolLearned}\subref*{fig:symbolLearned-1},
\texttt{Male} and \texttt{Female} are encoded as the shapes of astrological symbols.
In Figure \ref{fig:symbolLearned}\subref*{fig:symbolLearned-2},
\texttt{Male} and \texttt{Female} are encoded as the shapes of single characters.

\begin{CodeChunk}


\begin{flushleft}\includegraphics[width=0.4\linewidth]{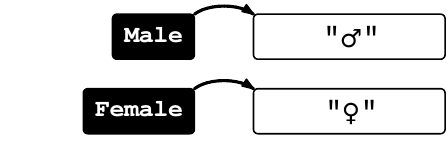} \includegraphics[width=0.4\linewidth]{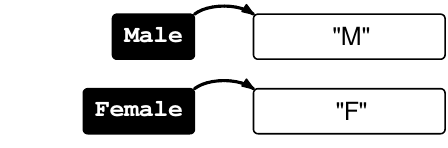} \end{flushleft}

\end{CodeChunk}

Again, these decodings are not as reliable as text words because
there are multiple possible decodings.

\begin{CodeChunk}


\begin{flushleft}\includegraphics[width=0.4\linewidth]{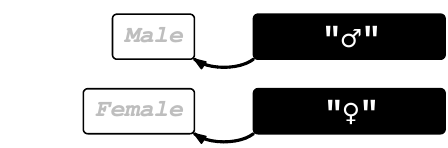} \includegraphics[width=0.4\linewidth]{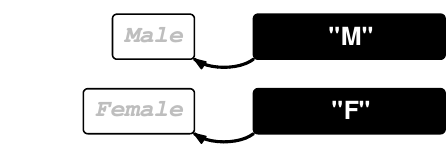} \end{flushleft}

\end{CodeChunk}

\begin{CodeChunk}


\begin{flushleft}\includegraphics[width=0.4\linewidth]{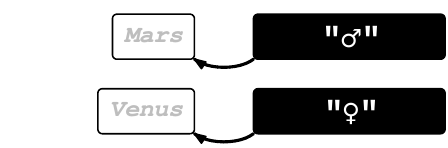} \includegraphics[width=0.4\linewidth]{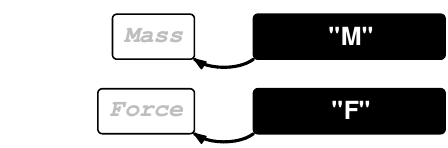} \end{flushleft}

\end{CodeChunk}

Another important caveat is that learned decodings often have
strong cultural or generational constraints.
For example, the male-blue and female-pink connection
mostly only applies to western cultures
and only since the mid-20th century
and is becoming anachronistic with greater recognition and tolerance of
non-binary identities.
This of coures also applies to the learned decoding of text shape.
All of the examples in this article are specific to an
English-reading audience.

Figure \ref{fig:rwcChernoff} shows
another example of data symbols with learned decodings.
In this case we have a set of Chernoff faces \citep{chernoff1973use}
representing the data from the 2023 Rugby World Cup.
For each country,
the number of points conceded is encoded as the spacing of the eyes,
the number of points scored is encoded as the curve of the
mouth and the number of tries scored is encoded as the angle of the eyebrows.
Angle and curvature, on their own, are considered to be poor visual features
for encoding data values, but the collection of simple lines
in each face produces an overall symbol that has a learned decoding
as a facial expression, and even a basic emotion.
In this case, we can easily identify a set of sad and worried teams
at top left (teams with fewer points scored, more points conceded, and
fewer tries)
and a another set of aggressively happy teams at
bottom right (teams with more points scored, fewer points conceded, and
more tries).
Again, the point is that non-text data symbols are capable of
learned decoding, but in this case, the result is arguably more
effective than any text representation could be.
We can read faces even more easily and readily than we
can read text. On the other hand, there is a limit to what
can be communicated through facial expressions alone.

\begin{CodeChunk}
\begin{figure}

{\centering \includegraphics[width=0.49\linewidth]{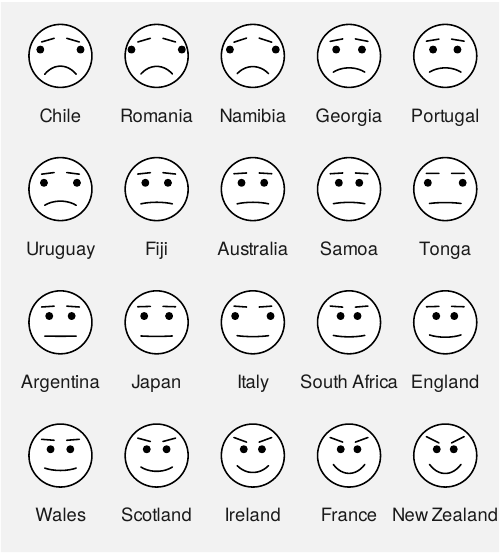} \includegraphics[width=0.49\linewidth]{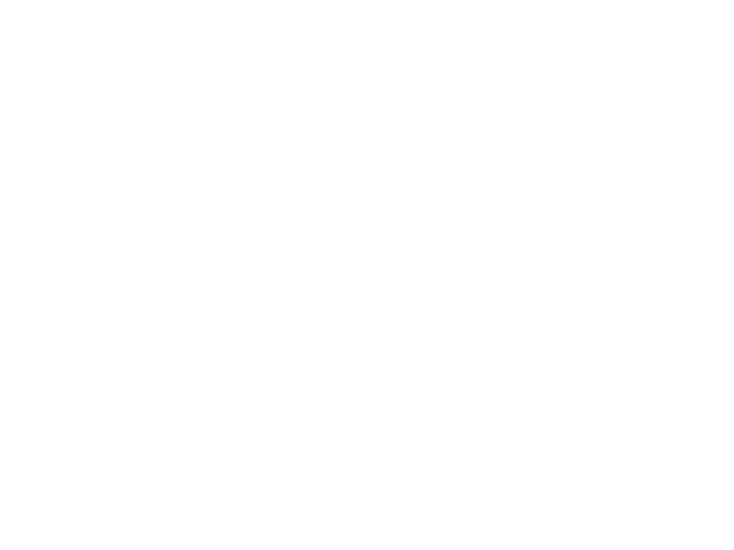} 

}

\caption[Chernoff faces showing the number of points scored (curve of the mouth), the number of tries (angle of the eye brows), and the number of points conceded (spacing of the eyes) for teams in the 2023 Rugby World Cup]{Chernoff faces showing the number of points scored (curve of the mouth), the number of tries (angle of the eye brows), and the number of points conceded (spacing of the eyes) for teams in the 2023 Rugby World Cup. We can decode clusters of similar teams based on similar facial expressions.}\label{fig:rwcChernoff}
\end{figure}
\end{CodeChunk}

In summary, the shape of text data symbols is not the only visual
feature that can have a learned decoding, although the
shape of text data symbols is in general the most
flexible and powerful combination of visual feature and data symbol.

\hypertarget{preattentive-decodings}{%
\subsection{Preattentive decodings}\label{preattentive-decodings}}

One downside to learned decodings is that they require learning.
Everyone has to learn to read.
This is in contrast with \emph{preattentive}
decodings that occur automatically, like
perception of position, length, area, and colour.
Preattentive decodings also occur very rapidly and in parallel,
whereas
learned decodings require more cognitive effort, take longer,
and occur serially \citep{ware2020information,brath2020visualizing}.

This is one reason why the bar plot
data visualisation in Figure \ref{fig:drama}
is much more effective than the 1,000 words.
The simple bar shapes, their colours and lengths are all
decoded rapidly and in parallel, whereas the 1,000 words must be read
serially and relatively slowly.
This is also the reason why most data visualisations
encode data values as the position, length, and colour of simple
geometric shapes.

However,
Figure \ref{fig:drama} shows that preattentive visual features, like colour,
still work for text. For example, we can rapidly identify the blue
elements in the plot, including the blue text in one of the titles,
without having
to read and comprehend the title.
The point is that it is the shape of text, and decoding semantic content
from that shape, that is slower and requires more cognitive effort.

In summary, if we encode data values as preattentive visual features of data
symbols, even text data symbols,
we get rapid decoding.
It only takes more time and effort
to get semantic content via learned decoding from the
shape of the text.
This is the primary achilles heel of text data symbols.

\hypertarget{ensemble-decodings}{%
\subsection{Ensemble decodings}\label{ensemble-decodings}}

In the scatter plot in Figure \ref{fig:rwc}\subref*{fig:rwc-1},
the quantitative data values are encoded as the
horizontal and vertical position of data points.
This allows us to decode a horizontal position to a number of points conceded
and a vertical position to a number of points scored,
but there is also another useful decoding available.

The combined positions in space of the data points in Figure
\ref{fig:rwc}\subref*{fig:rwc-1} allow us to identify separate clusters of
points and an overall trend or correlation between the
number of points conceded and the number of points scored
\citep{10.1167/16.5.11}.
For example, we can see that teams that scored more points tended to
concede fewer points, in general, and that there is
a group of three teams that performed worse on both measures
than all of the other teams.

This \emph{ensemble} decoding is possible when data values have been
encoded as the horizontal and vertical position of data points.
Another example of ensemble decoding is the ability to summarise
features of the bars in Figure \ref{fig:notext}.
We are easily able to perceive that the pink bars are on average lower
than the blue bars without having to focus on every individual blue or
pink bar (and similarly for the light green versus dark green bars).

However, ensemble decoding
is not possible for all visual features or for all data symbols.
In particular, we cannot rapidly summarise
semantic content from the shapes of a large number of text data symbols.
We have to read text one word at a time
\citep{brath2020visualizing}.

This is one reason why it is harder to extact trends from
Table \ref{tab:table} compared to the bar plot in Figure \ref{fig:drama}.
We can decode individual data values from a table of text, but we
are unable to visually summarise data values from a table of text.

Again, this limitation is specific to the learned decoding of text shape.
Ensemble decoding is still possible with preattentive visual features
of text data symbols.
For example,
a very basic visual trend is still visible in Table \ref{tab:table}
in the ``Male'' column because the length of the text data symbols
(the number of digits in each number)
provides a crude representation of the size of the numbers (and they
decrease from top to bottom within each year).\label{page:pretable}

Even the preattentive shape of text data symbols can still reveal
patterns and groupings. For example,
it is possible to perceive, without comprehending every individual data
value, that
the right-most digit of each number in Table \ref{tab:table}
is either a ``0'' or a ``5'' (the values have been rounded to the nearest
multiple of five).
The shapes at the ends of the numbers form identifiable groups
based on preattentive visual features, without requiring
comprehension of the numeric values.

In summary, large quantities of text data symbols are not effective
for representing large quantities of data values, if we encode
the data values as the shapes of the text.
On the other hand, if we encode data values as preattentive
features of text data values, patterns can still be perceived within
large blocks of text.

\hypertarget{combining-encodings}{%
\subsection{Combining encodings}\label{combining-encodings}}

Figure \ref{fig:mosaic}\subref*{fig:mosaic-1}
shows a mosaic plot of the number of male
and female offenders for different types of crime in 2024
(a different visualisation of the bottom-left bar plot in
Figure \ref{fig:drama}).
In this visualisation, the data symbol is a rectangle.
The proportion of crimes of each type are encoded as the heights of
the rectangles and the proportion of males and females, within each
type of crime, are encoded as the widths of the rectangles.

\begin{CodeChunk}
\begin{figure}

{\centering \subfloat[Data values are encoded as the widths and heights of the
rectangles, but we also perceive the area of the rectangles. We can decode the
crime type proportions from the heights of the rectangles and the offender sex
proportions (within each type of crime) from the widths of the rectangles \emph{and}
the overall proportion of each combination from the area of the rectangles.\label{fig:mosaic-1}]{\includegraphics[width=0.49\linewidth]{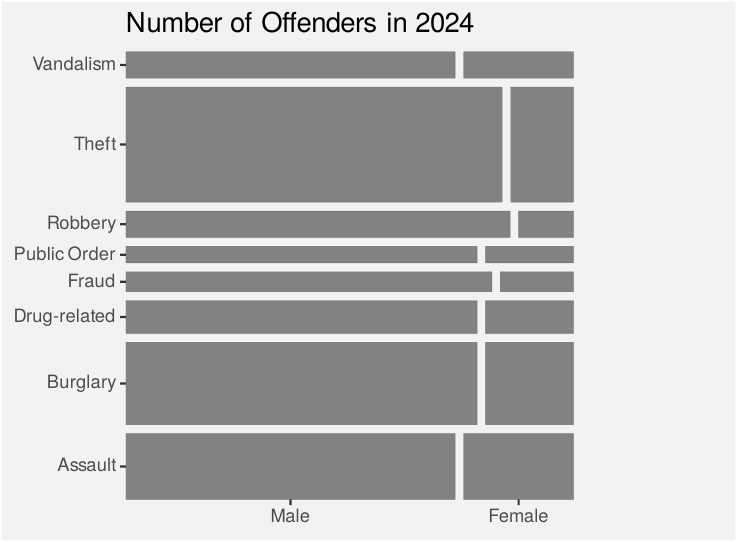} }\subfloat[The sex of the offender is redundantly encoded as the colour
\emph{and} the horizontal position of the rectangles; we can decode the sex of the
offender from the colour of the rectangles as well as from the horizontal
positions of the rectangles. The colours of the rectangles are perceived
independently of the rectangle sizes; the different colours of the rectangles
does not affect our ability to decode the different widths, heights, and areas
of the rectangles.\label{fig:mosaic-2}]{\includegraphics[width=0.49\linewidth]{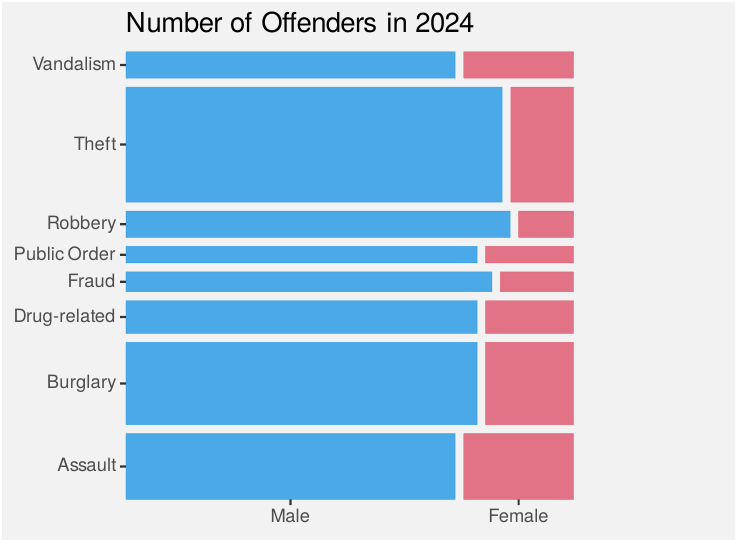} }

}

\caption[Mosaic plots of the number of male youth offenders and the number of female youth offenders for different types of crime in New Zealand in 2024]{Mosaic plots of the number of male youth offenders and the number of female youth offenders for different types of crime in New Zealand in 2024.}\label{fig:mosaic}
\end{figure}
\end{CodeChunk}

The interesting thing about these encodings is that they interact,
so that we can decode not just the widths and the heights of the
rectangles, but also their \emph{area} (which corresponds to the
overall proportion of offenders for each combination of sex
and crime type).

We can contrast this with the encodings within
Figure \ref{fig:mosaic}\subref*{fig:mosaic-2}, where
the sex of the offender is also encoded as the colour of the rectangles.
This colour encoding does not interact with the encodings of rectangle width
and height. The fact that a rectangle is blue or pink does not
affect our decoding of the rectangle width, height, or area.

Encodings that do not interact are typically more desirable because it makes
decoding easier. When encodings interact it can be more difficult
to decode the original data values
\citep{munzner2014visualization}.
The mosaic plot in
Figure \ref{fig:mosaic}\subref*{fig:mosaic-1}
is a relatively rare example
where visual features interact in a useful way.

Using text as a data symbol
does not necessarily change how visual features combine.
However, one consideration is whether text-specific visual
features interact.
\cite{brath2020visualizing} speculates that
some features, like capitalisation and font slant may interact
because they both affect the shape of the text, so using
more than one text-specific visual feature at once may not
be effective.

\hypertarget{redundant-encodings}{%
\subsection{Redundant encodings}\label{redundant-encodings}}

There is another way in which encodings are combined in
Figure \ref{fig:mosaic}\subref*{fig:mosaic-2}.
The sex of the offender is encoded in two ways:
as the horizontal position of the rectangles
and as the colour of the rectangles.
This sort of \emph{redundant encoding} can be useful to
strengthen the decoding of different qualitative data values
\citep{borkin7192646,VanderPlas03042017}.
Figure \ref{fig:mosaic}\subref*{fig:mosaic-1}
demonstrates that, without the additional colour encoding, it
is more difficult to identify which bar belongs to \texttt{Male}
versus \texttt{Female} (not impossible, just more difficult).

We can see that this is also true for redundant encoding with
text data symbols in
Figure \ref{fig:textaes}\subref*{fig:textaes-2}
versus
Figure \ref{fig:rwc}\subref*{fig:rwc-2}.
It is much easier to see the different groups (teams from different
hemispheres) in the former because the groups are encoded
using both the colour and the shape of the text data symbols,
while the groups are only encoded as the shape of the text in the latter.

In summary, using text as a data symbol
is just as effective as using geometric data symbols like bars, points,
and lines.
This is mostly because the effectiveness of encodings is dependent on
the choice of visual features rather than the choice of data symbol and
we can encode data values using standard visual features like
position and colour for text data symbols as well as for geometric
data symbols.
There are some visual features that are specific to text data symbols
and, while these provide greater flexibility for encoding data values
as text, text-specific visual features are not as effective as
the best standard visual features like position and colour.

There are two ways in which text data symbols surpass other
options like bars and points.
Encoding data values
as the shape of text data symbols,
as words or numbers,
provides the greatest capacity for representing different categories and
provides the most accurate way to decode a single data value from
a data visualisation.
On the other hand, the accuracy and convenience of text declines
if we are required to compare two or more data values.
Another downside is that we cannot decode large quantities of words
or numbers as rapidly as we can decode large quantities of bar lengths
or coloured bars.
Furthermore, we cannot calculate summaries like trends and clusters
from large quantities of numbers, which is something that we can do with large
quantities of geometric data symbols.

\hypertarget{sec:phrases}{%
\section{Phrases and sentences}\label{sec:phrases}}

In Sections \ref{sec:chars}, \ref{sec:words}, and \ref{sec:numbers},
we only considered encoding data values as characters, words, or
numbers,
although this has managed to cover text labels on axes
and legends as well
as text data symbols within the plot region.
In this section we consider larger text labels, which are more common
in titles on
axes and in the overall plot title or caption.

For example, in Figure \ref{fig:rwc}\subref*{fig:rwc-1}
the axis titles describe the
variable that is encoded on each axis and the plot title describes
the overall data topic.
We can still think of these titles
as encodings, but encodings of metadata rather than encodings of data values.

\begin{CodeChunk}


\begin{flushleft}\includegraphics[width=0.5\linewidth]{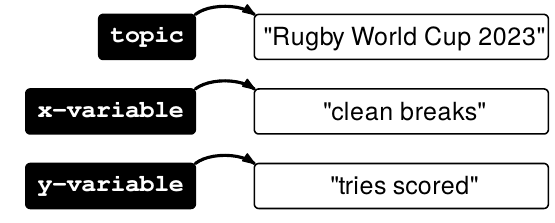} \end{flushleft}

\end{CodeChunk}

One characteristic of the values that are being encoded is that they
contain more complex information, such as the fact that the data
relates to the 2023 Rugby World Cup.
Text shape is the most effective way to encode this sort of complex information.
Not only is it possible to decode text to obtain semantic content, but
there is no limit to the complexity of that content.

As we saw in Section \ref{learned},
it is not impossible to express more complex information without text,
but non-verbal alternatives are typically limited and less effective.
For example, Figure \ref{fig:rwcTitle}\subref*{fig:rwcTitle-1}
shows a version of
Figure \ref{fig:rwc}\subref*{fig:rwc-1}, with the title
drawn using non-text symbols.
The concept of the Rugby World Cup is conveyed using a silhouette
of the Web Ellis Cup, which is the trophy that is presented to the
winner, plus ten dots to indicate that this was the tenth event of its
kind. This ``title'' is clearly more difficult to decode than the
original text title.
A better alternative might be to use the official Rugby World Cup 2023
logo, although that would still only resonate with avid fans
of the tournament (and in any case the logo copyright precludes
that option).

Another important point is that it is still possible to take
advantage of
preattentive decoding in larger text elements.
For example, in Figure \ref{fig:drama}, the main title
is bold. We can view this as an encoding of the relative importance
of the main title as the font face of the title text.
By comparison, the text on the axes is smaller;
the lower importance of these text elements is encoded as the size
of the text.

A more subtle point is that we can also make use of
preattentive decoding within just
a subset of the text. For example, only specific words
are coloured in the two secondary titles in Figure \ref{fig:drama}.
The data value \texttt{Males} is encoded as the colour of just the
word ``Males'' within the secondary title on the left.
Furthermore, the similarity of the colours within the titles
and the colours of the bars creates visual groups that connect
the learned decoding of the text semantics to the corresponding
bars.

A final point is that the positions of the secondary titles, the proximity
of the titles to the columns of bars (and their common left alignment),
helps to create visual groups between each secondary title
and its associated column of bars \citep{Todorovic:2008}.

\hypertarget{annotations}{%
\subsection{Annotations}\label{annotations}}

Another common example of a larger text element within a data visualisation
is an annotation that identifies significant values.
Again, encoding the information as the shape of text is a good choice
because of the ability to convey more complex information.

Figure \ref{fig:rwcTitle}\subref*{fig:rwcTitle-2} shows
an example of a more verbose annotation that
explains the Rugby World Cup result to disillusioned New Zealanders.
This is an example of complex information that would be
impossible to represent using non-text symbols.

\begin{CodeChunk}
\begin{figure}

{\centering \subfloat[The title consists of non-text symbols.  It is difficult to decode the semantic content of this title.\label{fig:rwcTitle-1}]{\includegraphics[width=0.49\linewidth]{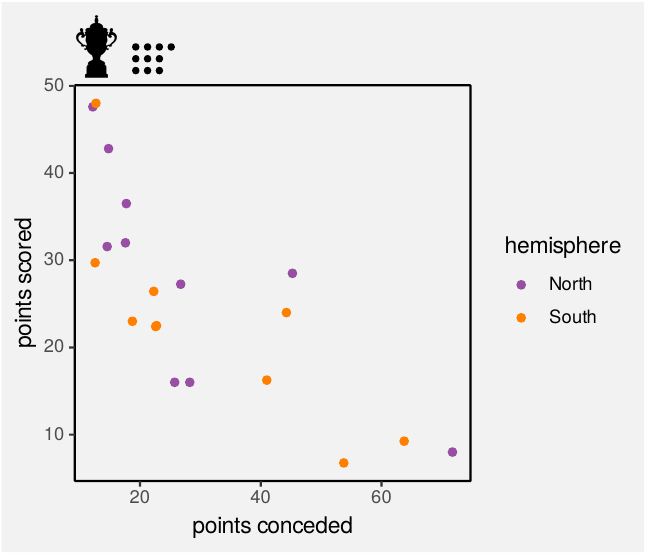} }\subfloat[An annotation provides additional context for the data from the top two teams.\label{fig:rwcTitle-2}]{\includegraphics[width=0.49\linewidth]{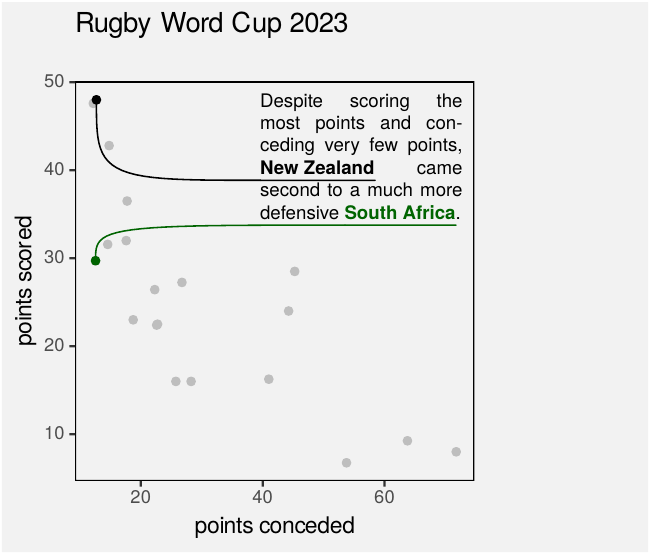} }

}

\caption[Scatter plots of the number of points scored and the number of points conceded (both are per-game averages) for teams in the 2023 Rugby World Cup]{Scatter plots of the number of points scored and the number of points conceded (both are per-game averages) for teams in the 2023 Rugby World Cup.}\label{fig:rwcTitle}
\end{figure}
\end{CodeChunk}

Again, there are also preattentive decodings at play in this example.
There are lines and colours connecting specific words within the annotation
to the relevant data points within the scatter plot.
This creates visual groups that associate the learned decoding of the
text to the corresponding data points
\citep{Todorovic:2008}.

\hypertarget{sec:blocks}{%
\subsection{Larger blocks of text}\label{sec:blocks}}

Paragraphs and sections and chapters of text are not
typically considered data visualisations.
However, they are still a form of visual communication and indeed
symbolic communication, so are not entirely unrelated to data visualisation.

The unquestionable advantage of a large block of text is the
ability to communicate arbitrarily complex and sophisticated messages.
For example, in Figure \ref{fig:drama} the text block contains not just
lists of raw data values, and simple summary features of the data, but also
discussions of whether the data features generalise to other contexts
and what the implications are for efforts to reduce youth crime.
The downside, as we have noted, is that a large block of text has to
be read sequentially in order to extract that more complex information.
However, it is also possible to convey simpler messages within
larger blocks of text.

Even in the main text of this article, there are demonstrations
of preattentive decodings. For example, the distinction between
section headings and normal paragraphs is encoded as the
weight and size of the text: headings are bolder and larger.
The importance of particular terms are also encoded as \emph{italic} text within
normal paragraphs.
These are essentially encodings of different categories.

In summary, the use of larger phrases and sentences for titles,
captions, and annotations in a data visualisation
makes sense because text shape is the most
effective way, if not the only way, to encode complex concepts
and information.
At the same time, we can still effectively encode basic data values,
such as categories, and less complex
information, such as relative importance,
using colour, size, and text-specific visual features that only
require preattentive decoding.
Although, as we consider larger amounts of text,
we depart further from what is typically considered data visualisation,
it is useful to think of a continuum rather than a dichotomy.
Text symbols can be used to a greater or lesser extent to communicate
information and preattentive versus learned decodings can be
employed in all possible combinations.

\hypertarget{sec:mem}{%
\section{Engagement and memorability}\label{sec:mem}}

In Section \ref{sec:effectiveness} we explored the effectiveness of
text as a representation of data values.
That section focused on how well data values can be decoded
from text data symbols.
However, the effectiveness of a data visualisation can also be
measured in other ways \citep{kosara-mackinlay-2013,bertini-2020}.
For example, \cite{few2004chartjunk} suggests three ways that
a data visualisation might be effective, beyond just accurate
representations of the data:

\begin{compactenum}
\item \say{by engaging the interest of the reader (i.e., getting them to read 
      the content).}
\item \say{by drawing the reader's attention to particular items that merit 
      emphasis.}
\item \say{by making the message more memorable.}
\end{compactenum}

There is evidence that text plays an important role
in all of these aspects.
The text elements within a data visualisation,
especially larger elements such as titles and annotations,
are paid the most attention
\citep{borkin7192646,DVS,stokes2022textoftenbetterthemes}.
Text has a large influence on the message that is
taken from a data visualisation, possibly larger
than any message that is conveyed by other data symbols
\citep{kong-2018}.
There is also evidence that text increases the memorability
of a data visualisation \citep{borkin7192646}.

In summary, larger text components are not only the appropriate way to encode
complex information, but they play an important role in
attracting attention to a data visualisation and in ensuring
that the information that we decode from a data visualisation
is not immediately forgotten.

\hypertarget{sec:tables}{%
\section{Tables}\label{sec:tables}}

In Section \ref{sec:intro} we lumped Table \ref{tab:table} in with
the block of text in Figure \ref{fig:drama} as largely text-based
presentations of the New Zealand youth crime data and contrasted that
with the bar plot data visualisation in Figure \ref{fig:drama}.

We have since argued against this strict dichotomy and suggested
that text is just another symbol that we can use to represent information.
We can now revisit Table \ref{tab:table}, treat it as less of a straw
man contrast with bar plots, and more seriously consider the
effectiveness of the table in terms of the encoding and decoding of data values.

In Table \ref{tab:table}, all raw data values are encoded as the shape
of text data symbols. This is very effective for decoding individual
data values, which is the main strength of the table.
The downside is that it is slow, it requires effort to decode
multiple data values, and we are unable to perform ensemble decoding
of trends and correlations, based on decoding data values from the
text shapes.

Variable names and group names are encoded as text symbols
in the row of column headings. Again, this is very effective
because of the learned decoding of text, especially when we
need to convey more complex concepts, like sex or a range of ages.

However, in addition to the encoding of information as the shape of text,
there is also information encoded in other, preattentive visual features
of the text.
In particular, the position of the text in Table \ref{tab:table}
encodes several important aspects of the data.
The horizontal position of the text encodes the variable or group
for each data value. The greater horizontal spacing between columns
compared to the vertical spacing between rows produces clear
columns of values \citep{Todorovic:2008}.
This is assisted, more subtly, by the justification of the text
(left or right) within each column.
The vertical position of the text encodes the type of offence and the
year for each data value and the vertical positioning of
rows is clustered to create a strong visual group for each year.
In other words, there is information encoded in the preattentive
features of the text that allows us to perceive structure within the
table.

As noted previously (page \pageref{page:pretable}),
the lengths of the numbers also permit preattentive decoding
of crude trends across type of offence, within years.
We can easily perceive that the upper crime types within each
year are more common and that pattern persists across years.
And we can do that without having to read and comprehend
each individual number.

We also noted previously, that the values in Table \ref{tab:table}
are rounded to the nearest 5, which is one example of what are
considered good practice guidelines for laying out tables
\citep{j.2333-8504.1992.tb01443.x,Schwabish_2020}.
The value of this guideline is preattentive good Gestalt
\citep{Todorovic:2008}, rather than learned decoding
of the precise data values.

The positions of the text symbols in Table \ref{tab:table},
particularly the top row and the first
two columns,
act very much like the axes on a data visualisation (with
cartesian coordinates).
The positions of these labels allows us to associate a count
with a particular sex or age and type of offence and year.
Figure \ref{fig:tablePlot}\subref*{fig:tablePlot-1}
emphasises this similarity by drawing a plot with text data symbols
to represent the counts for different combinations sex, year, and type
of offence.
Figure \ref{fig:tablePlot}\subref*{fig:tablePlot-2}
goes one step
further and encodes the magnitude of the counts as the size of the text
data symbols.
These two figures demonstrate that there is a clear overlap
between tables and what would normally be considered data visualisations
and that there is a continuum rather than a dichotomy.

\begin{CodeChunk}
\begin{figure}

{\centering \subfloat[The data symbols are numbers. This data visualisation
bears a strong resemblance to Table \ref{tab:table}.\label{fig:tablePlot-1}]{\includegraphics[width=0.49\linewidth]{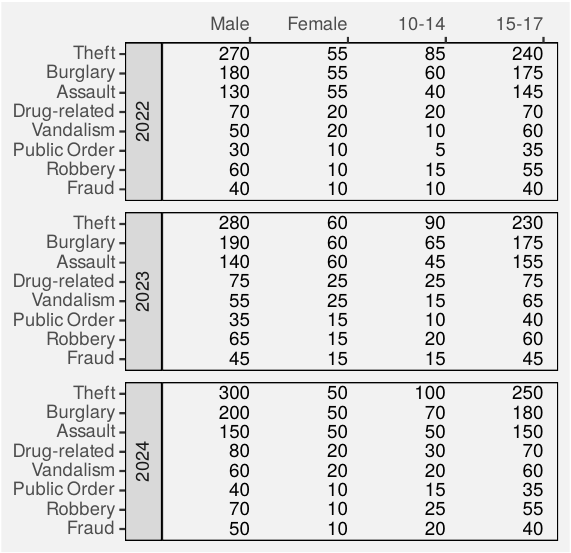} }\subfloat[A facetted plot with numbers as the data symbols and the
magnitude of the counts encoded as the size of the data symbols. This is more
obviously a data visualisation, but it still strongly resembles Table
\ref{tab:table}.\label{fig:tablePlot-2}]{\includegraphics[width=0.49\linewidth]{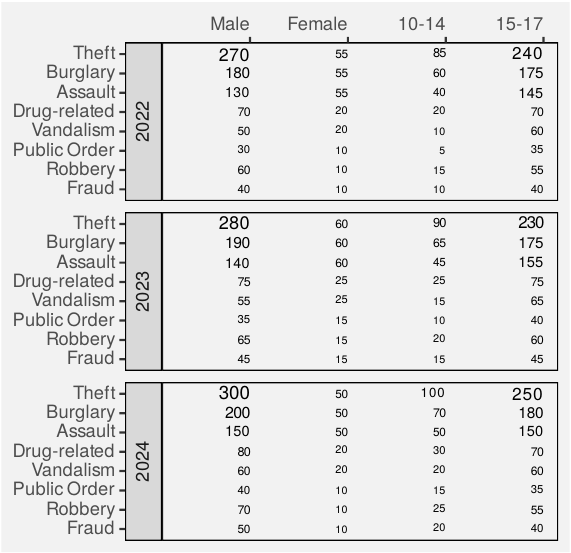} }

}

\caption[Data visualisations (facetted plots) of the number of youth offenders in New Zealand from 2022 to 2024, broken down by sex, age group, and type of offence]{Data visualisations (facetted plots) of the number of youth offenders in New Zealand from 2022 to 2024, broken down by sex, age group, and type of offence.}\label{fig:tablePlot}
\end{figure}
\end{CodeChunk}

\hypertarget{sec:summary}{%
\section{Conclusion}\label{sec:summary}}

We began with a complaint:
that there was no one-stop shop for an explanation of how text works
in a data visualisation.
This article attempts to plug that gap.

The main contribution is a framework that conceives of text as just
another option for visually presenting information.
This framework is summarised below:

\begin{itemize}
\item[]  A data visualisation is a communication with the viewer.  Information is
  encoded as the visual features of symbols and the task for the viewer is to
  decode the visual features of the symbols to retrieve the information.

  The text within a data visualisation can be understood as just another symbol
  that is used to encode information, just like a bar in a bar plot or a data
  point in a scatter plot.  For example, axis and legend labels are just data
  values encoded as text.

  Information can be encoded as preattentive visual features of text, like
  colour, position, and size, just like for other data symbols.  This allows
  rapid, parallel, and even ensemble decoding of information from text.

  Text also has its own text-specific preattentive visual features, like weight
  and slant.  This provides an opportunity to encode more data variables within
  a text data symbol, though these text-specific visual features have relatively
  low capacity and relatively low accuracy.

  The text elements within a data visualisation are subject to the same Gestalt
  principles as other visual elements.  Proximity and similarity are important
  tools for grouping text labels with non-text symbols.  This is essential to
  the effectiveness of axes and legends, but can also be used elsewhere, for
  example, in titles.

  Encoding information as the shape of text---as words and numbers---is
  especially powerful because of the learned decoding of information from words.
  This allows very expressive encoding and very precise decoding.  The downside
  is that this decoding is slower and serial and therefore less effective for
  large amounts of text.  Furthermore, the decoding of text shape is vulnerable
  to cultural contexts just like any other learned decoding.

  Text is typically the \emph{only} encoding in a data visualisation that has a
  learned decoding.  In other words, text is typically the only element within a
  data visualisation that has semantic content.  That is why it is so essential
  to any data visualisation.  Text is the best way to present exact decoding in
  axes and legends and the only way to convey complex information in titles and
  annotations.

  On the other hand, much more time and effort is required to decode the
  semantic content of a large amount of text.  This is why data visualisations
  usually contain large numbers of non-text data symbols with data encoded as
  preattentive visual features of those data symbols.
\end{itemize}

Overall, the emphasis in this article is less on handling text as
a special type of data to be visualised, and more on including text as
another option for encoding any type of data. Furthermore,
we view all text on a data visualisation as an encoding of data or
information, rather than distinguishing between text that
``supports'' and text that encodes.
As a consequence, we gain greater insight into how text functions
as part of a data visualisation.
We see that text is very effective for decoding specific data values,
which is essential to the functioning of axes and legends.
We see that text is less effective for ensemble decoding, which is
where simpler geometric data symbols excel.
However, there is no blanket prohibition on using large
amounts of text in a data visualisation.
This can still makes sense if we are relying on
ensemble perception of preattentive features, like the colour or size of text.
We just cannot rely on ensemble perception
from the learned decoding of text shape.

This framework is a synthesis and a simplification of
many ideas and results from the data visualisation literature.
One limitation of a simplified framework is its simplicity.
Treating phrases and sentences and paragraphs of text simply as
visual encodings
of information that are to be decoded from the shapes of their
component characters severely understates the
mental processes that are involved.
For example,
although text must at first follow the same visual pathway within the brain
as any other visual input, we know that
there are processing areas
that deal specifically with language \citep{ware2020information}.
Furthermore, the comprehension of language involves cognitive processes
that are much more complex than the perception of
preattentive visual features
and text plays a prominant role in directing where we look,
not just what we see
\citep{10.1145/3593580}.
Nevertheless, the hope is that there is value in
a framework that de-emphasises some of those details in order
to provide a straightforward way to reason about the use of text
within a data visualisation.

\hypertarget{acknowledgements}{%
\section{Acknowledgements}\label{acknowledgements}}

The data set and the block of text in
Figure \ref{fig:drama} were generated with the assistance of
\href{https://copilot.microsoft.com/}{Microsoft Copilot},
based on information from the
\href{https://www.youthcourt.govt.nz/youth-justice/youth-justice-statistics/}{Youth Court of New Zealand},
\href{https://www.justice.govt.nz/justice-sector-policy/research-data/justice-statistics/youth-justice-indicators/}{New Zealand Ministry of Justice}, and
\href{https://www.stats.govt.nz/topics/crime-and-justice}{Stats NZ}.
The counts should not be treated as official statistics.

The data set for the 2023 Rugby World Cup was obtained from
\href{https://en.wikipedia.org/wiki/2023_Rugby_World_Cup_statistics}{Wikipedia}.

The male and female symbols in Figure \ref{fig:symbolLearned} are
from the \href{https://fonts.google.com/specimen/Martian+Mono}{Martian Mono font}.

The paragraphs of text in Figure \ref{fig:drama} was typeset
with \href{https://fonts.google.com/specimen/DM+Sans}{DM Sans Extra Light}.

All data visualisations were generated in \proglang{R} \citep{Rlang}
using the following packages:
\pkg{grid},
\pkg{gridBezier} \citep{pkg:gridBezier},
\pkg{gridtext} \citep{pkg:gridtext},
\pkg{gridGeometry} \citep{pkg:gridGeometry},
\pkg{ggplot2} \citep{ggplot2},
\pkg{gggrid} \citep{pkg:gggrid},
\pkg{ggtext} \citep{pkg:ggtext},
\pkg{ggforce} \citep{pkg:ggforce},
\pkg{ggChernoff} \citep{pkg:ggChernoff},
\pkg{ggmosaic} \citep{pkg:ggmosaic},
\pkg{vwline} \citep{pkg:vwline},
\pkg{xdvir} \citep{pkg:xdvir}, and
\pkg{png} \citep{pkg:png}.

\hypertarget{appendix-appendix}{%
\appendix}

\hypertarget{app:a}{%
\section{Text in the data visualisation literature}\label{app:a}}

In Section \ref{sec:intro}, we attempted to demonstrate
that there was a need for a more consistent and coherent treatment
of the role of text in data visualisation.
In order to make the demonstration straightforward,
we leaned into examples that supported that thesis.
This appendix acknowledges and admits to some of the omissions and
simplifications and other transgressions
that were committed in Section \ref{sec:intro}.

The framework proposed in this article draws together a large
amount of research and insight from other sources.
For example, we have explored the effectiveness of text data symbols
by considering what we already know about the effectiveness of
non-text data symbols from works like \cite{munzner2014visualization}.
Furthermore, 
some of the main points from the summary in Section \ref{sec:summary}
can also be found in, for example, \cite[p. 327]{ware2020information}:

\begin{quotation}
\say{When a large number of data points must be represented in a visualization,
use symbols instead of words or pictorial icons.}
\end{quotation}

\begin{quotation}
\say{Use words directly on the chart where the number of symbolic objects
in each category is relatively few and where space is available.}
\end{quotation}

However, the information within \cite{ware2020information} is disconnected
from discussions of the effectiveness of other representations of data.
This is part of our argument that the information about text is
more fragmented and less coherent than infmrmation about other 
representaions of data.

In Section \ref{sec:intro} we argued that there was a paucity
of information about employing text as data symbols, but that
information is not completely
absent from the literature. For example, Chapter 10
of \cite{schwabish2021better}
contains multiple examples of text-based presentations of data,
particularly for when the data itself is text.
However, the use of text is often separated from the use
of other types of data symbols.
This article attempts to integrate text as just another way
to encode information within a data visualisation;
as just another data symbol.

Several articles have explicitly addressed the encoding of data values
as preattentive visual features of text, reaching back to
Bertin \citep{Brath04032019}.
\cite{bertininfluence} even referred to text as
``the 4th symbol type'' after points, lines, and area.
However, much of this work has occurred in the context of maps
and did not explore the role of text
in data visualisations in general.

There have been examples of blurring the boundaries between tables and
plots.
For example,
a forest plot \citep{Lewis1479} is a presentation
of meta-analysis results that combines tabular values
with line and point data symbols.
Another example is the \pkg{sparkTable} package for R \citep{RJ-2015-003}, which
allows sparklines \citep{tufteSparklinesHistory} to be included within tables.
This article just goes further and proposes that these boundaries be torn down.

The integration of text is also addressed comprehensively
in \cite{brath2020visualizing}.
Brath acknowledges
that there are some clear reasons why text may not
generally be thought of as a valid component of data visualisation
\citep[p. 28]{brath2020visualizing}:

\begin{quotation} 
\say{Unlike preattentive visual attributes, text must be
read. Reading is linear: one-word-at-a-time. Preattention is rapid parallel
perceptual processing. Compared to preattention, reading text is slow to
comprehend. Further, preattention can immediately detect a red dot amongst many
gray dots but finding one word among many is not preattentive. Following this
path of logic, text does not fit the criteria for visualization.}
\end{quotation}

However, Brath points out that data values can be encoded as preattentive
visual features of text, like colour and size.
Furthermore, Brath identifies text-specific preattentive visual features
of text.

This article follows a very similar line of reasoning.
We have acknowledged that text is different from simple geometric data symbols
because it involves a learned decoding, which is slow and serial,
rather than preattentive decoding, which occurs rapidly and in parallel.
On the other hand,
we have identified that text still has preattentive features and
that there are non-text data symbols with
learned decodings as well.
As a consequence of the overlaps between text and non-text
data symbols, they should be considered comparable rather than incommensurable.

\begin{figure}
\includegraphics{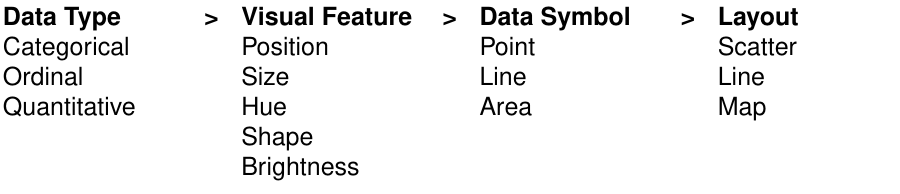}
\caption{\label{fig:bertin}Bertin's visual encoding pipeline.  
Adapted from Figure 2.8 of \cite{brath2020visualizing}.}
\end{figure}

Despite these similarities, we have approached
the integration of text in data visualisations in at least two significantly
different ways compared to \cite{brath2020visualizing}.
The first difference involves the integration of text into what Brath refers to
as the ``visual encoding pipeline''
for transforming data values into a data visualistion.
Figure \ref{fig:bertin} shows a pipeline based on \cite{bertin1983semiology},
which only considers non-text data symbols.
There are different types of data, which can be encoded using different
visual features of different data symbols, and drawn within different layouts.

\begin{figure}
\includegraphics{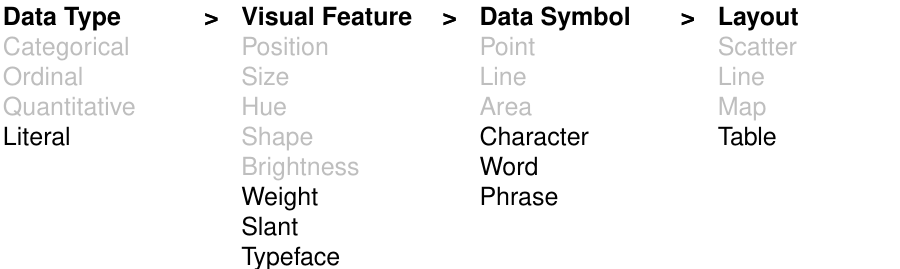}
\caption{\label{fig:brath}Brath's extension of Bertin's 
visual encoding pipeline to include text.  The old components are grey
and the new components are black.
Adapted from Figure 2.13 of \cite{brath2020visualizing}.}
\end{figure}

Figure \ref{fig:brath} shows how \cite{brath2020visualizing} extends
the pipeline to integrate text data symbols.
There are additional text-based data symbols and
there are text-specific visual features, and
tables layouts are included, consistent with this article.
In addition,
there is a new type of data called ``literal text'',
which corresponds to when the data values are text
(e.g., survey responses).

\begin{figure}
\includegraphics{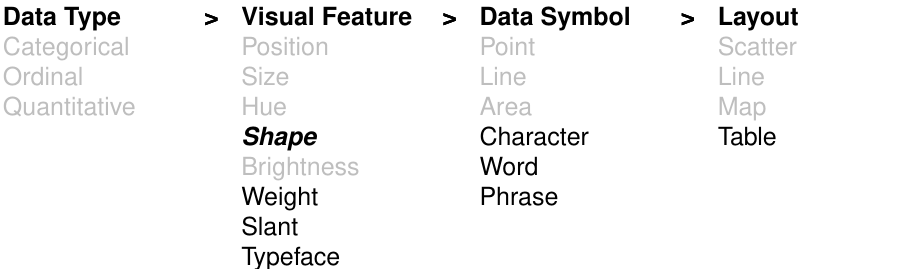}
\caption{\label{fig:murrell}This article's extension of Bertin's 
visual encoding pipeline to include text.   The old components are grey,
new components are black, and the modified component is black, bold, 
and italic.}
\end{figure}

This article does not invoke a new data type in order to integrate text
data symbols. Text is considered to be just another way to encode
the existing types of data.
The emphasis is instead on the special nature of
the \emph{shape} of text symbols, which enables learned decoding of
exact data values.
We might include a
new data type in the pipeline, but that would be higher-level
and more complex types of data such as information, knowledge,
and understanding \citep{ackoff1989data},
which can only really be encoded and decoded
via text.
Figure \ref{fig:murrell} shows how we might extend the pipeline
according to the description in this article.

Another difference is that Brath
still makes a distinction between the elements of a data visualisation
that represent data values and the ``supporting elements''
(largely text) such as titles, axis labels, and annotations
\cite[Figure 2.7]{brath2020visualizing}.
In this article, we simply
treat all elements of a data visualisation as symbols that
encode some sort of data and information.

\bibliography{murrell-dataverb.bib}

\end{document}